\documentclass[traditabstract]{aa}

\usepackage{graphicx}
\usepackage{txfonts}

\usepackage{epsf}
\usepackage{rotating}
\usepackage{graphics}
\usepackage{lscape}
\usepackage{pdflscape}

\def \MJ{M$_{\mathrm{Jup}}$}
\def \RJ{R$_{\mathrm{Jup}}$}

\def \kms{km\,s$^{-1}$}
\def \ms{m\,s$^{-1}$}
\def \1s{$1\,\sigma$}

\def \t0{T$_0$}

\newcommand{\kepler}{\emph{Kepler}}
\newcommand{\corot}{\emph{CoRoT}}
\newcommand{\Mjup}{M$_\mathrm{Jup}$}
\newcommand{\Rjup}{R$_\mathrm{Jup}$}

\def\Msun{\hbox{$\mathrm{M}_{\odot}$}}            
\def\Rsun{\hbox{$\mathrm{R}_{\odot}$}}
\def\met{[Fe/H]}

\newcommand{\teff}{\mbox{$T_{\rm eff}$}}
\newcommand{\logg}{\mbox{$\log g_*$}}
\newcommand{\vsini}{\mbox{$v \sin i_*$}}

\begin{document}
      
\title{Characterization of the four new transiting planets \\KOI-188b, KOI-195b, KOI-192b, and KOI-830b}           

\author{
G.~H\'ebrard\inst{1,2}
\and A.~Santerne \inst{3,4,5} 
\and G.~Montagnier \inst{1,2}
\and G.~Bruno \inst{3}
\and M.~Deleuil \inst{3}
\and M.~Havel \inst{4}
\and J.-M.~Almenara \inst{3}
\and C.~Damiani \inst{3}
\and S.\,C.\,C.~Barros \inst{3}
\and A.\,S.~Bonomo \inst{6}
\and F.~Bouchy \inst{3}
\and R.\,F. D\'{\i}az \inst{3,7}
\and C.~Moutou \inst{3}
}

\institute{
Institut d'astrophysique de Paris, UMR7095 CNRS, Universit\'e Pierre \& Marie Curie, 
98bis boulevard Arago, 75014 Paris, France 
\email{hebrard@iap.fr}
\and
Observatoire de Haute-Provence, CNRS/OAMP, 04870 Saint-Michel-l'Observatoire, France
\and
Aix Marseille Universit\'e, CNRS, LAM (Laboratoire d'Astrophysique de Marseille) UMR 7326, 13388 Marseille, France
\and
Centro de Astrof{\'\i}sica, Universidade do Porto, Rua das Estrelas, 4150-762 Porto, Portugal
\and
Instituto de Astrof\'isica e Ci\^{e}ncias do Espa\c co, Universidade do Porto, CAUP, Rua das Estrelas, 4150-762 Porto, Portugal
\and
INAF - Osservatorio Astrofisico di Torino, Via Osservatorio 20, 10025 Pino Torinese, Italy
\and
Observatoire astronomique de l'Universit\'e de Gen\`eve, 51 chemin des Maillettes, 1290 Versoix, Switzerland
}

\date{Received TBC; accepted TBC}
      
\abstract{
The characterization of four new transiting extrasolar planets is presented here.
KOI-188b and KOI-195b are bloated hot Saturns, with orbital periods of 3.8 and 3.2~days, and masses 
of 0.25 and 0.34~\MJ. They are located in the low-mass range of known transiting, 
giant planets.
KOI-192b has a similar mass (0.29~\MJ) but a longer orbital period of 10.3~days. This places it 
in a domain where only a few planets are known.
KOI-830b, finally, with a mass of 1.27~\MJ\ and a period of 3.5~days, is a typical hot Jupiter.
The four planets have radii of 0.98, 1.09, 1.2, and 1.08~\RJ, respectively.
We detected no significant eccentricity in any of the systems, 
while the accuracy of our data does not rule out possible moderate eccentricities.
The four objects were first identified by the \kepler\ Team as promising candidates from the photometry 
of the \kepler\ satellite. We  establish here their planetary nature thanks to the radial velocity follow-up we secured 
with the  HARPS-N spectrograph at the \textit{Telescopio Nazionale Galileo}.
The combined analyses of the  datasets allow us to fully characterize the four planetary systems. 
These new objects increase the number of well-characterized exoplanets for statistics, and provide new targets 
for individual follow-up studies.
The pre-screening we performed with the SOPHIE spectrograph at the \textit{Observatoire de Haute-Provence} 
as part of that study also allowed us to conclude that a fifth candidate, 
KOI-219.01, is not a planet but is instead a false 
positive\thanks{Table~\ref{PriorTable} is available only in electronic 
              form at the CDS via anonymous ftp to cdsarc.u-strasbg.fr 
              (130.79.128.5) or via http://cdsweb.u-strasbg.fr/cgi-bin/qcat?J/A+A/.
              CDS also includes the radial velocities given in Tables~\ref{table_rv_sophie} and~\ref{table_rv}.}. 
}
  

\authorrunning{H\'ebrard et al.}
\titlerunning{Characterization of the four new transiting planets KOI-188b, KOI-195b, KOI-192b, and KOI-830b}

\maketitle

\section{Introduction}

Today, more than 3800 transiting planetary candidates have been identified from the data 
of the \kepler\ satellite. They have been obtained from the analyses of the light curves 
of the 156\,000 stars with magnitudes $9<V<16$ continuously observed by  \kepler\ 
from May 2009 to May 2013 with a high photometric accuracy.
These candidates are designated as KOIs (Kepler Objects of Interest) in the successive announcements  
by the \kepler\ Team (e.g., Borucki et al.~\cite{borucki11a}, \cite{borucki11b}; Batalha et al.~\cite{batalha12};
Burke et al.~\cite{burke13}).
Whereas several stellar configurations can mimic planetary transits 
(e.g.,~Almenara et al.~\cite{almenara09}), it has been 
argued that contrary to ground-based or \corot\ photometric surveys, such false positives 
are rare among \kepler\ candidates and most of them should actually 
be planets (e.g.,~Morton \&\ Johnson~\cite{morton11}). This is particularly the case 
for multiple-planet candidates (e.g., Lissauer et al.~\cite{lissauer12},~\cite{lissauer14}). 
On the other hand, the proportion of false positives is expected to be significant 
among single, close-in, giant exoplanet KOI candidates, with a false positive 
rate of $34.8\pm6.5$\%\ measured by Santerne et al.~(\cite{santerne12}), 
whereas Fressin et al.~(\cite{fressin13}) predicted it to be $29.3\pm3.1$\%\ from statistical 
arguments (both false positive rates are computed for periods shorter than 25~days). 
Such false positive rates are lower than those of ground-based and 
\corot\ photometric surveys but they remain significant.
Some individual cases of KOI false positives have been presented, e.g., by 
Bouchy et al.~(\cite{bouchy11}), Col\`on et al.~(\cite{colon12}), 
D\'{\i}az et al.~(\cite{diaz13}), or Moutou et al.~(\cite{moutou13}).

Follow-up analyses and/or observations are thus mandatory in order to identify which KOIs
are actually planets, and which are not. Such identifications are essential in order to construct 
a sample of exoplanets free of  false positives for unbiased statistical studies.
This is also important for individual analyses of particular objects, as transiting 
planets allow numerous studies including atmospheric absorber detections or 
obliquity measurements. 
Radial velocity is a particularly rich follow-up observation method as it allows numerous 
stellar scenarios to be distinguished from actual planets among photometric candidates.
For the identified planets, it also allows their masses and the eccentricity of their orbits to be 
measured. The planetary radii and masses are provided by transit light curves and 
radial velocities, respectively; the joint use of the two methods gives access to the 
planetary densities and deep characterizations.

\begin{table*}[ht!]
\centering
\caption{IDs, coordinates, and magnitudes of the planet-host stars.}            
\begin{tabular}{lcccc}       
\hline                
\kepler\ Object of Interest  & KOI-188 & KOI-195 & KOI-192 & KOI-830 \\
\kepler\ exoplanet catalog & Kepler-425 & Kepler-426 & Kepler-427 & Kepler-428 \\
\hline 
\kepler\ Input Catalog 	& KIC5357901 			& KIC11502867 		& KIC7950644 			& KIC5358624 			\\
USNO-A2 ID  	& 1275-11218459		& 1350-10395462		& 1275-10946568		& 1275-11249110		\\
2MASS ID   	& 19212592+4034038  	& 19174431+4928242 	& 19130109+4342175	& 19221961+4034386	\\
\hline            
RA (J2000)   	& 19:21:25.92		& 19:17:44.31 	& 19:13:01.10 		& 19:22:19.62 \\   
DEC (J2000) 	& +40:34:03.86  	& +49:28:24.24 &  +43:42:17.53	& +40:34:38.64 \\  
\hline
\kepler\ magnitude $K_{\rm p}$			& 14.74 &  14.84 & 14.22 &  15.22 \\
Johnson-$V$	&	$14.97 \pm 0.04$	&	$15.073 \pm 0.013$	&	$14.42 \pm 0.02$	&	$15.405 \pm 0.016$	\\ 
Johnson-$B$	&	$16.047 \pm 0.016$	&	$15.736 \pm 0.021$	&	$15.13 \pm 0.03$	&	$16.584 \pm 0.046$	\\ 
SDSS-$G$	&	$15.465 \pm 0.014$	&	$15.293 \pm 0.019$	&	$14.65 \pm 0.03$	&	$15.931 \pm 0.024$	\\ 
SDSS-$R$	&	$14.73 \pm 0.07$	&	$14.863 \pm 0.035$	&	$14.31 \pm 0.07$	&	$15.101 \pm 0.037$	\\   
SDSS-$I$	&	$14.605 \pm 0.22$	&	$14.735 \pm 0.029$	&	$14.013 \pm 0.019$	&	$14.84 \pm 0.07$	\\     
2MASS-$J$  	&	$13.376 \pm 0.021$	&	$13.624 \pm 0.026$	&	$13.161 \pm 0.023$	&	$13.767 \pm 0.024$ 	\\ 
2MASS-$H$ 	&	$12.931 \pm 0.020$	&	$13.230 \pm 0.033$	&	$12.814 \pm 0.022$	&	$13.295 \pm 0.023$ 	\\
2MASS-$K_s$	&	$12.806 \pm 0.027$	&	$13.212 \pm 0.033$	&	$12.777 \pm 0.024$	&	$13.234 \pm 0.033$ 	\\
WISE-$W1$	&	$12.791 \pm 0.025$	&	$13.200 \pm 0.024$	&	$12.731 \pm 0.023$	&	$13.142 \pm 0.025$ 	\\     
WISE-$W2$	&	$12.864 \pm 0.027$	&	$13.252 \pm 0.029$	&	$12.782 \pm 0.024$	&	$13.292 \pm 0.031$ 	\\
WISE-$W3$	&	$12.662$		&	$13.089$		&	$12.72 \pm 0.32$	&	 $13.222$ 		\\                              
\hline
\end{tabular}
\label{startable_KOI}      
\end{table*}

Since 2010 we have been conducting radial-velocity follow-up of KOIs with the SOPHIE 
spectrograph at the \textit{Observatoire de Haute-Provence} (OHP, France) to characterize 
\kepler\ candidates. We mainly focus on the brightest stars (\kepler\ magnitude 
$K_p < 14.7$) harboring close-in giant planet candidates. 
This has allowed us to 
identify and characterize several new transiting planets 
(Santerne et al.~\cite{santerne11a}, \cite{santerne11b}; Bonomo et al.~\cite{bonomo12}; 
Deleuil et al.~\cite{deleuil14}),
as well as more massive companions and false positives 
(Ehrenreich et al.~\cite{ehrenreich11a}; Bouchy et al.~\cite{bouchy11}; Santerne et 
al.~\cite{santerne12}; D\'{\i}az et al.~\cite{diaz13}; Moutou et al.~\cite{moutou13}).
We are extending our on-going SOPHIE program on KOIs characterization 
with the HARPS-N spectrograph at the \textit{Telescopio Nazionale Galileo}
(TNG, La Palma, Spain), taking advantage of its higher radial-velocity accuracy for 
fainter targets.
Our observation strategy with HARPS-N complements SOPHIE observations in three ways.
First, we use HARPS-N to follow KOIs for which our SOPHIE data provides only a 
detection hint or an upper limit on the planetary mass, but with a precision preventing 
firm conclusion and accurate characterization. 
Second, we use HARPS-N to follow KOIs fainter than the limit $K_p = 14.7$ 
adopted in the SOPHIE sample. 
Third, we also use HARPS-N to follow KOIs with shallower transits than those observed 
with SOPHIE.
As part of that HARPS-N program, we previously reported (H\'ebrard et al.~\cite{hebrard13a})
the characterization of the two new 
transiting planets KOI-200b and KOI-889b (afterwards named Kepler-74b and Kepler-75b
by the \kepler\ Team); these planets  complied with the conditions of detection hint and 
faintness, respectively.
Both are giant planets on 7.3- and 8.9-day orbits, but whereas KOI-200b has 
a mass of $0.68 \pm 0.09$~\MJ, KOI-889b is a massive planet of $9.9 \pm 0.5$~\MJ.
These two planets were among the first ones to be detected with HARPS-N.
HARPS-N has also been used to measure the mass 
of the Earth-sized planets Kepler-78b (Pepe et al.~\cite{pepe13}; 
see also Sanchis-Ojeda et al.~\cite{sanchis13} 
and Howard et al.~\cite{howard13})
and Kepler-10b,c (Dumusque et al.~\cite{dumusque14}), 
and also to study obliquities in planetary
systems  (Covino et al.~\cite{covino13}; 
Esposito et al.~\cite{esposito14};  
Santerne et al.~\cite{santerne14}; 
Bourrier \&\ H\'ebrard~\cite{bourrier14}; Lopez-Morales et al.~\cite{lopez14}), 
to show that the metal-poor star HIP\,11952 does not harbor giant planets 
(Desidera et al.~\cite{desidera13}), 
and to study the planetary system around XO-2S (Desidera et al.~\cite{desidera14}).

Here we present the characterization of four new transiting planets with 
HARPS-N, namely KOI-188b, KOI-195b, KOI-192b, and KOI-830b. 
The IDs and coordinates of the four planet-host stars are given in Table~\ref{startable_KOI}, 
which also presents their spectral energy distributions (SEDs) from magnitudes at different wavelengths.
KOI-192 was in the SOPHIE 
sample, but its nature was not established and only an upper mass of $0.6$~\MJ\ could be put on the 
transiting object if actually it was a planet (Santerne et al.~\cite{santerne12}). 
The other three  objects are fainter than the magnitude limit of the SOPHIE sample. 
We also report here the identification of KOI-219.01 as a false positive, which is in our third 
sample which includes shallower transit candidates.

We describe the photometric and spectroscopic observations of the targets in 
Sect.~\ref{sect_observations}, the analysis of the  datasets 
in Sect.~\ref{sect_analysis}, and the results and their discussion in Sect.~\ref{sect_disc}.

\begin{figure*}[]
\begin{center}
\begin{tabular}{cc}
\hspace{-0.3cm}
\includegraphics[width=\columnwidth]{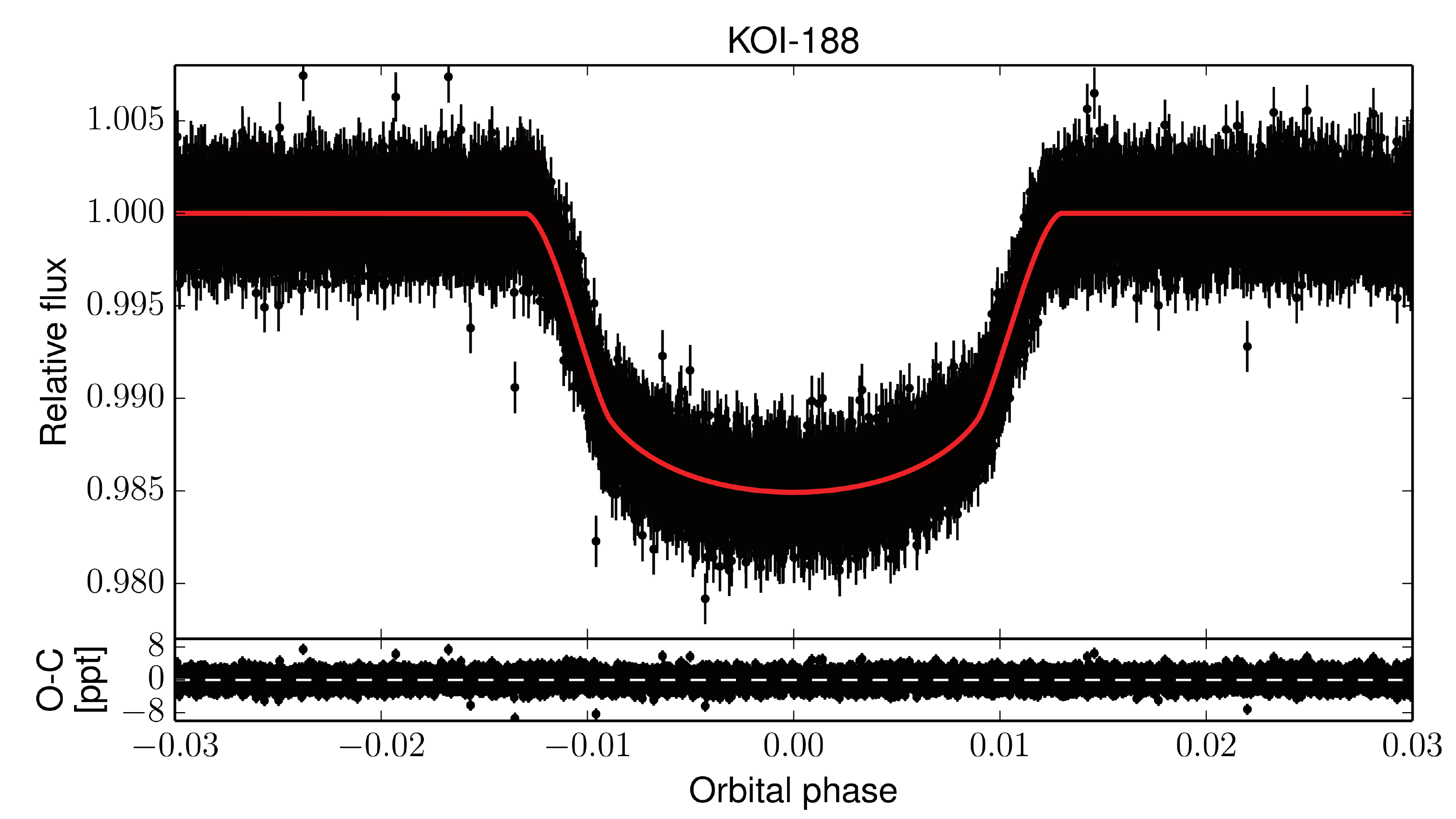} & \includegraphics[width=\columnwidth]{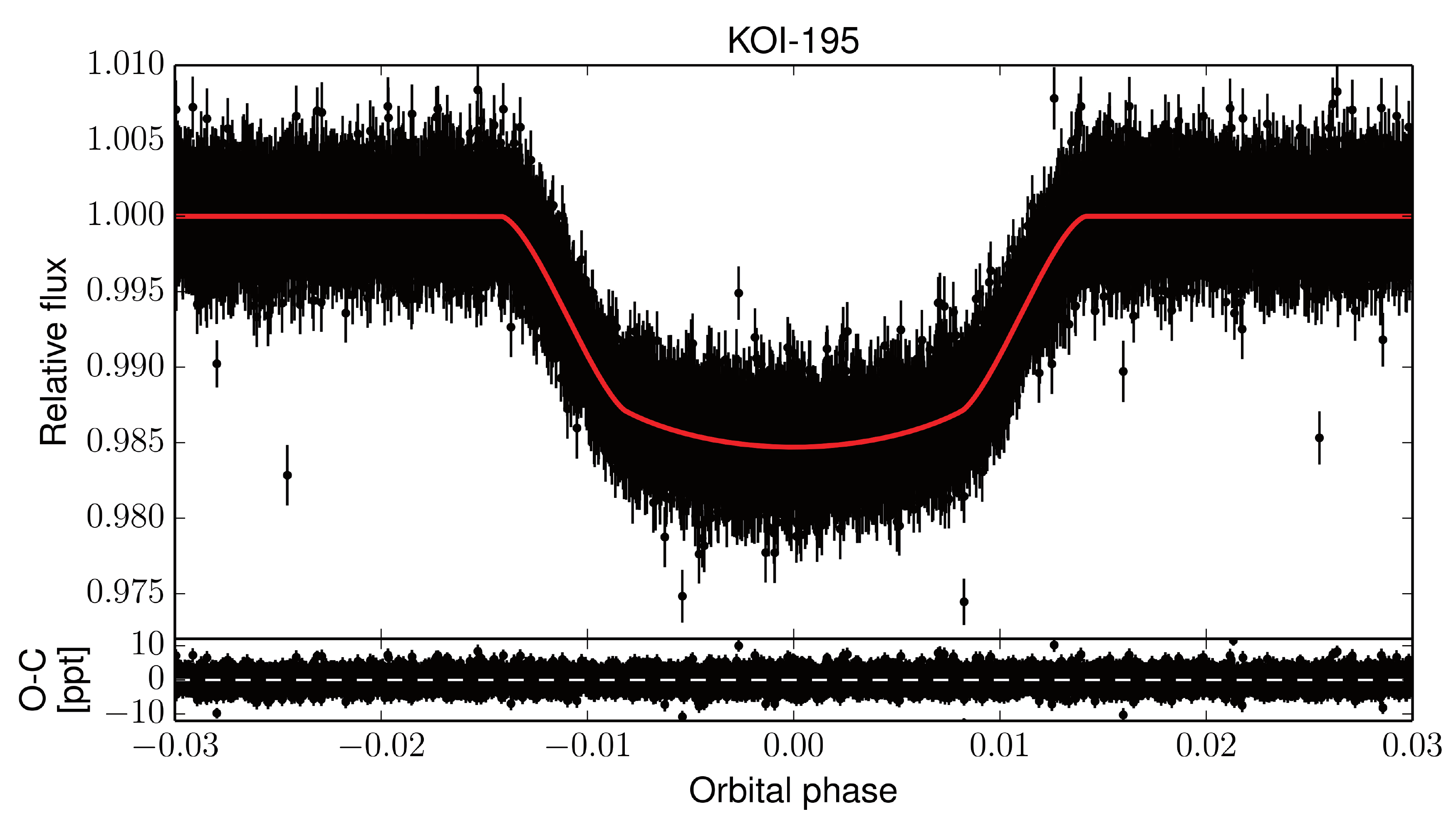}\\
\hspace{-0.3cm}
\includegraphics[width=\columnwidth]{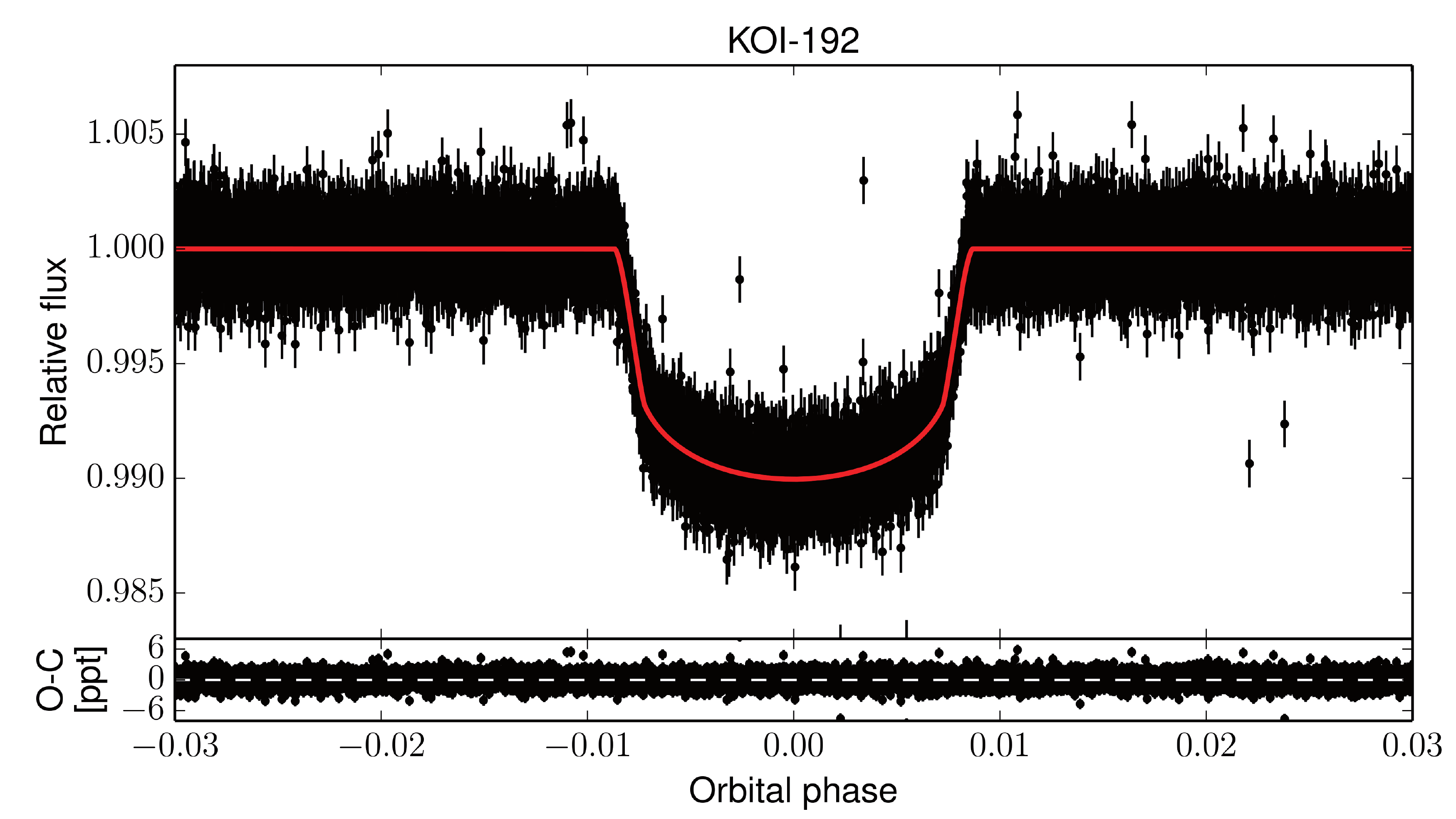} & \includegraphics[width=\columnwidth]{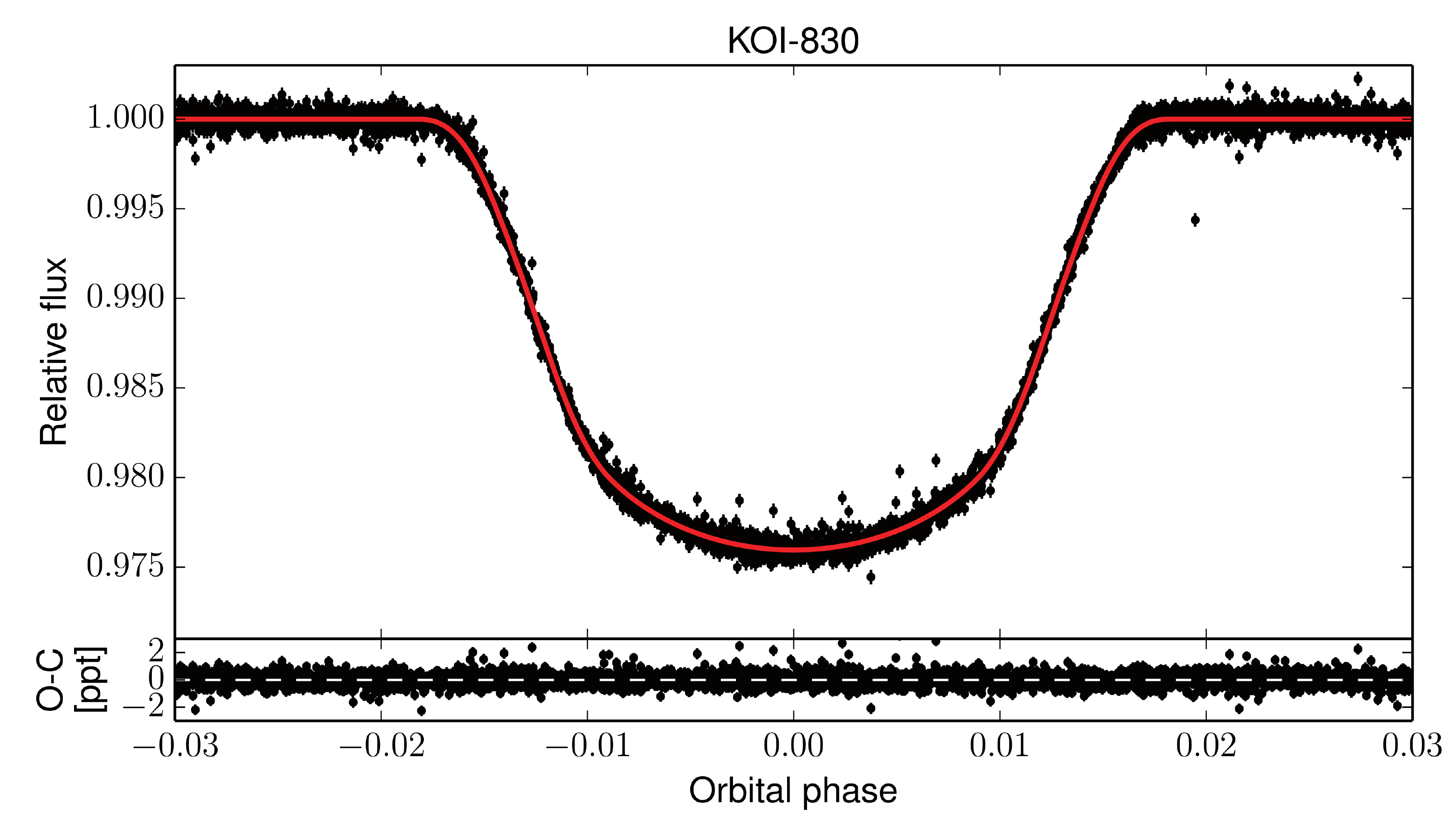}\\
\end{tabular}
\caption{\kepler\ photometric data of the four planet-host stars. 
On each plot the upper panel shows 
the phase-folded light-curve (black dots with 1-$\sigma$ error bars) over-plotted with the best model 
(red line; see Sect.~\ref{sect_parameter_system}), 
and the lower panel shows the residuals. The models fit together the photometric data, the radial velocities, and the 
spectral energy distribution 
(Table~\ref{table_rv} and Fig.~\ref{bestmodel_RV}). 
The parameters of the fits are reported in Tables~\ref{posterior_1} and~\ref{posterior_2}.
Only \kepler\ short-cadence data are plotted here for KOI-188, KOI-195, and KOI-192, 
but long-cadence data are also used in the analysis of these three systems.
No short-cadence data are available for KOI-830, whose long-cadence data are plotted here.
}
\label{bestmodel_LC}
\end{center}
\end{figure*}

\section{Observations and data reduction}
\label{sect_observations}

\subsection{Photometric detection with \kepler}
\label{kepphot}

The five targets were observed by \kepler\ since the beginning 
of the mission in May 2009, and were identified by Borucki et 
al.~(\cite{borucki11a}, \cite{borucki11b}) and Batalha et al.~(\cite{batalha12})
as hosting single periodic transits with periods of a few days and 
depths characteristic of giant planets.
No transits with different periods were detected in 
any of the light curves, so there are no signs of multiple transiting systems.

The \kepler\ photometry was acquired in long-cadence data (LC, one point 
per 29.42~minutes). Short-cadence data (SC, one point per 58~seconds) are also 
available for KOI-188, KOI-195, and KOI-192.
Both LC and SC were used in our present analyses.
We used the light-curve of quarters Q1 to Q17
reduced by the Photometric Analysis 
\kepler\ pipeline that accounts for barycentric, cosmic ray, background, and so-called
argabrightening corrections (Jenkins et al.~\cite{jenkins10}), publicly available from the 
MAST archive (http://archive.stsci.edu/kepler).
The  light curves clearly present transits with depths of about 1\,\%.

They also show smooth variations of the flux which  could be due, at least partially, 
to inhomogeneities of the stellar surfaces (spots, plages, etc) modulated with the rotation of the stars.
The stronger variations are seen on KOI-188 with an amplitude up to 2500\,ppm in normalized flux.
The variations are weaker for the three other targets, with amplitudes up to 700, 
500, and 400\,ppm for KOI-195, KOI-192, and KOI-830, respectively.
We used Lomb-Scargle periodograms and autocorrelations of the light curves to attempt 
detections of stellar rotation periods. In the case of KOI-188, we obtained different rotation periods 
in a range between 10 and 40 days, depending on the section of the light curve we used.
We thus conclude that the signal is not significant enough to allow a reliable stellar rotation 
period to be derived. The signal is even less significant for the three other targets.
We note that Walkowicz \& Basri~(\cite{walkowicz13}) reported $P_{\rm rot} = 35.00\pm8.32$~days
from the \kepler\ light curve of KOI-188,~the large uncertainty probably reflecting the weaknesses of the 
signal. They do not report any detection for the  other three~targets.

Before modeling the transits, we normalized fragments of the light curves by fitting an 
out-of-transit parabola, 
first without accounting for contamination. Since the \kepler\  spacecraft rotates four 
times a year, the crowding values are different between seasons. We thus produced four 
crowding-uncorrected de-trended light curves for each target, one per season. This  
allowed us to account for differential crowding values, noises, and out-of-transit fluxes 
in the transit modeling and in the final 
error~budget (see Sect.~\ref{sect_parameter_system}). 
Figure~\ref{bestmodel_LC} shows the corresponding phase-folded light curves, 
once normalized and corrected for~crowding.

\subsection{Radial velocities}
\label{sect_RV}

\subsubsection{Pre-screening with SOPHIE}
\label{sect_RV_sophie}

Although we observed KOI-830 only with HARPS-N, we first observed the other four  targets  
with SOPHIE, the fiber-fed, echelle spectrograph mounted at the focus of the 1.93~m OHP telescope 
and dedicated to high-precision radial velocity measurements (Perruchot et al.~\cite{perruchot08}; 
Bouchy et al.~\cite{bouchy09},~\cite{bouchy13}). 
Even if the precision is lower than the HARPS-N measurements for such faint targets, SOPHIE 
measurements can reveal large radial velocity variations corresponding to transiting stars, 
brown dwarfs, or massive planets, which would not require HARPS-N to be characterized 
(see, e.g., Ehrenreich et al.~\cite{ehrenreich11a}; D\'{\i}az et al.~\cite{diaz13}; 
Moutou et al.~\cite{moutou13}; H\'ebrard et al.~\cite{hebrard13a}). 
The observations were made in high-efficiency mode (resolution power 
$\lambda/\Delta\lambda=40\,000$) using the slow read-out mode of the detector.
The exposure times ranged between 20 and 55 minutes, allowing radial velocity accuracies 
between $\pm20$ and $\pm50$~\ms\ to be~reached (see Table~\ref{table_rv_sophie}). 

The three observations of KOI-219 we secured with SOPHIE over a week in summer 2013 
revealed a binary star. Depending on the observation, we detected two sets of barely resolved 
spectral lines separated by a few \kms, or a unique set of spectral lines that we interpreted as 
the superposition of the two sets being located at a similar radial velocity at that time. 
The corresponding radial velocities and bisector spans than we measured on these spectra 
show a clear correlation (Fig.~\ref{fig_biss}), which 
means that the observed apparent radial velocity variations are mainly due to line profile 
variations rather than to Doppler shifts of the whole spectra.
That candidate is thus likely to be an unresolved eclipsing stellar binary which is diluted in an associated 
triple system or in a foreground/background star. The full resolution of the binary scenario would
require more observations and analyses, beyond the main scope of the present paper.
Here we conclude that this transiting candidate is not a planet but a false positive, so we did not 
pursue the observation of KOI-219 with the higher precision of HARPS-N. 
This illustrates the benefits that could be obtained for the follow-up of transiting planet candidates
from coordinated observations secured with two spectrographs with different sensitivities, precisions, 
and accessibilities as SOPHIE and HARPS-N.
Similar cases of diluted eclipsing binaries revealed with SOPHIE among the KOIs were presented 
by Santerne et al.~(\cite{santerne12}).

\begin{figure*}[]
\begin{center}
\begin{tabular}{c}
\includegraphics[width=18cm]{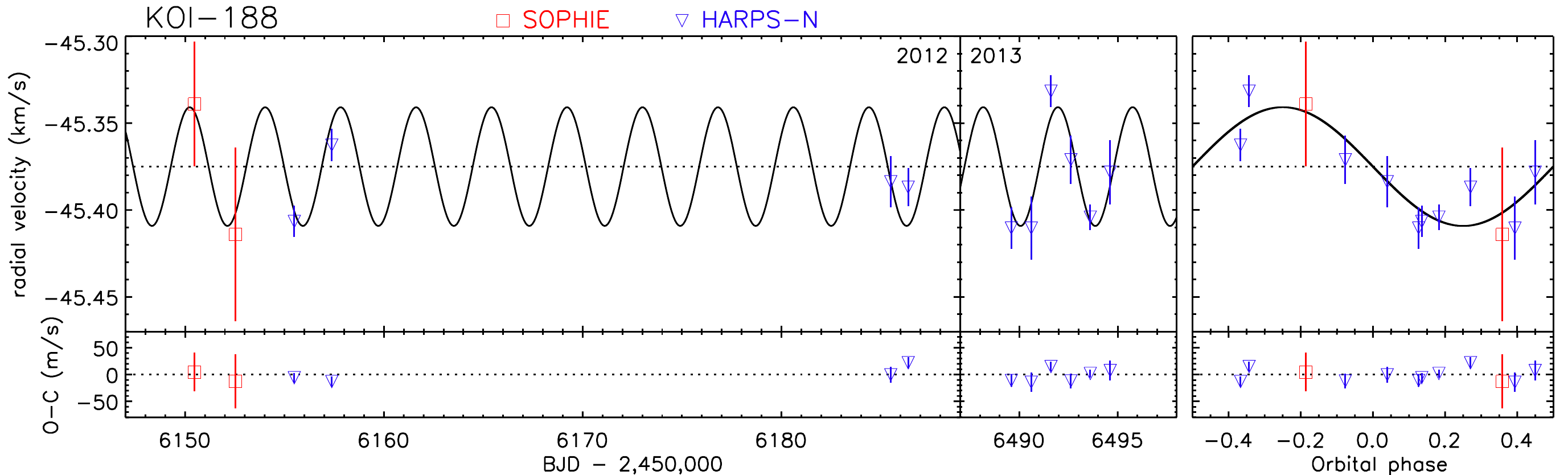} \\
\includegraphics[width=18cm]{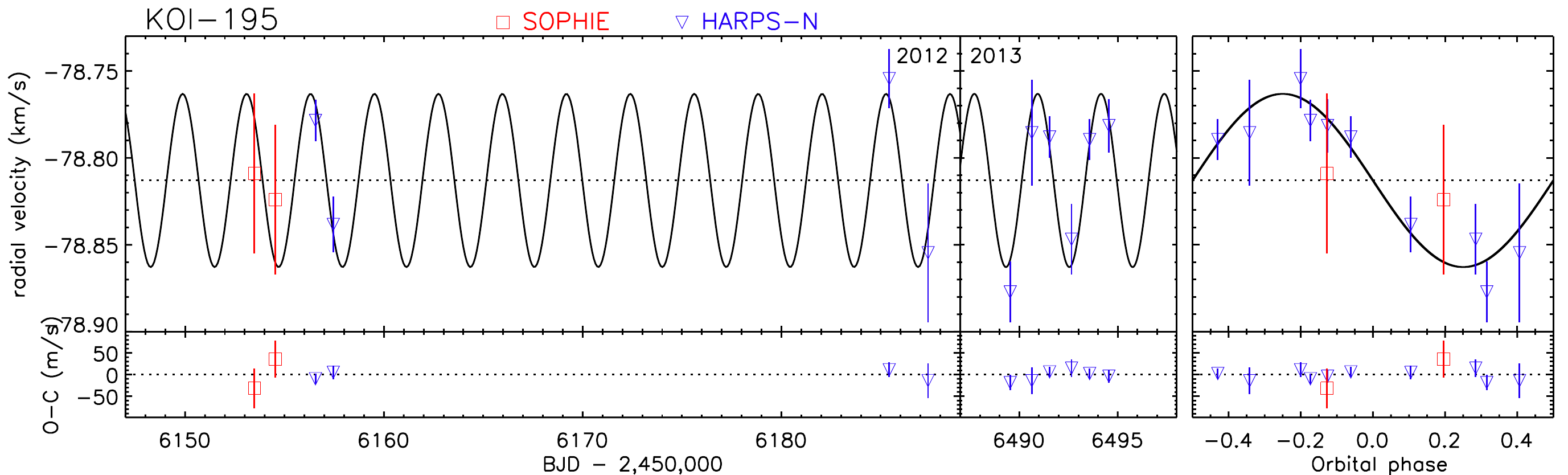} \\
\includegraphics[width=18cm]{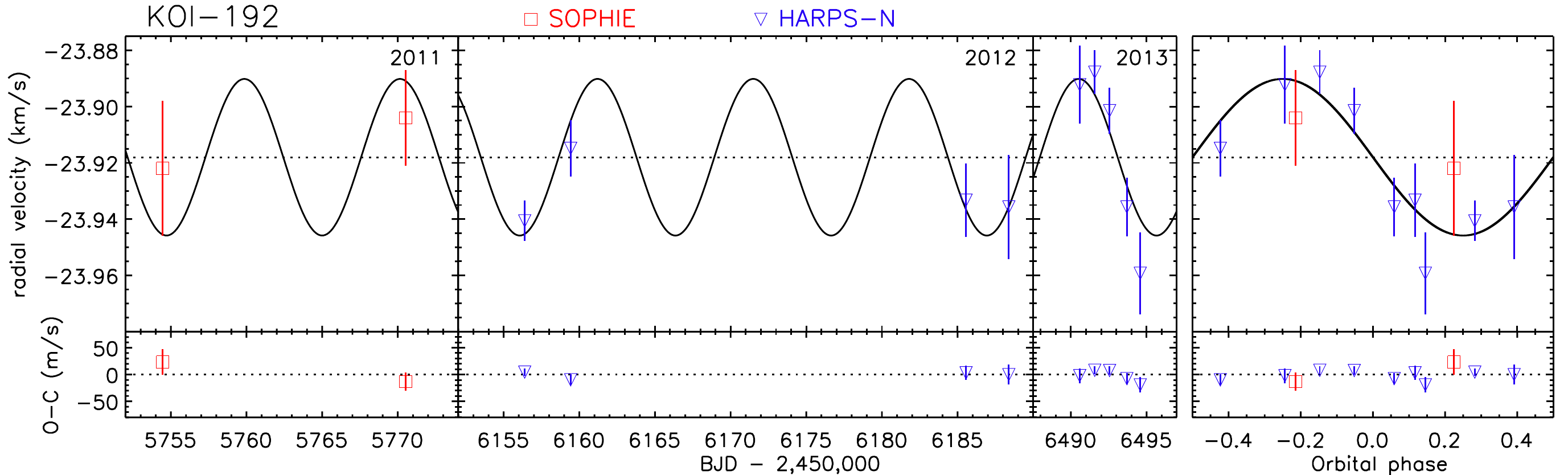} \\
\includegraphics[width=18cm]{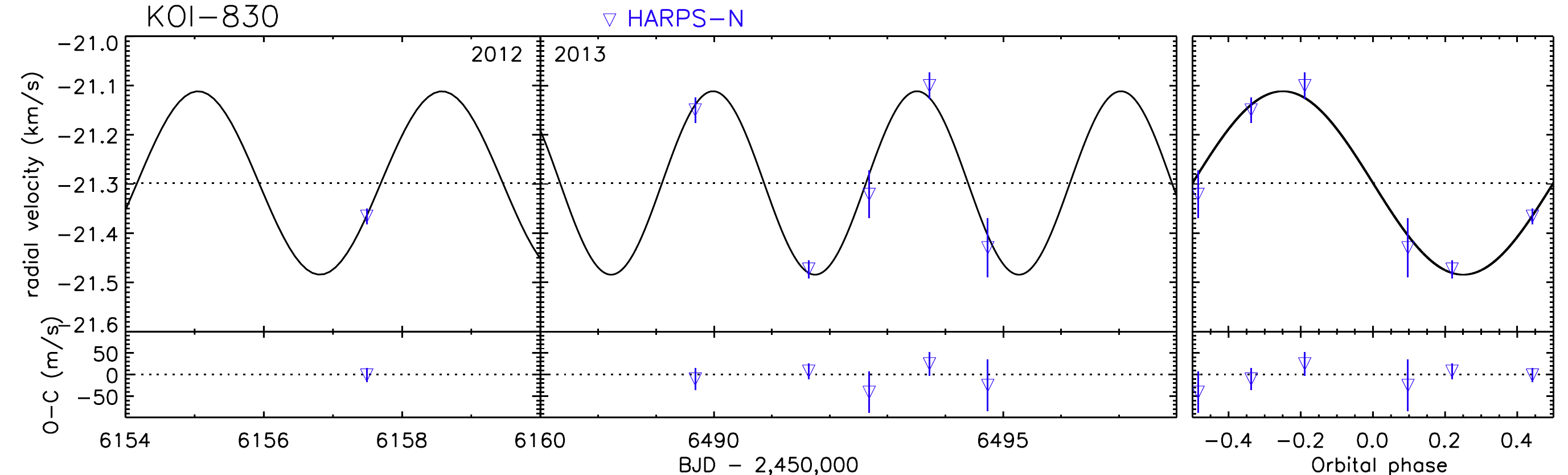} \\
\end{tabular}
\caption{Radial velocities of the four planet-host stars, from HARPS-N (blue) and SOPHIE (red), and 1-$\sigma$ 
error bars.
On each of the  four plots, the left panel shows the radial velocities as a function of time,
the right panel shows the phase-folded data,
the upper panel shows the data over-plotted with 
the best circular model (black line), and the lower panel shows the residuals.
The parameters of the fits are given in Tables~\ref{posterior_1}~and~\ref{posterior_2}.}
\label{bestmodel_RV}
\end{center}
\end{figure*}

\begin{figure*}[]
 \centering
\includegraphics[scale=0.5]{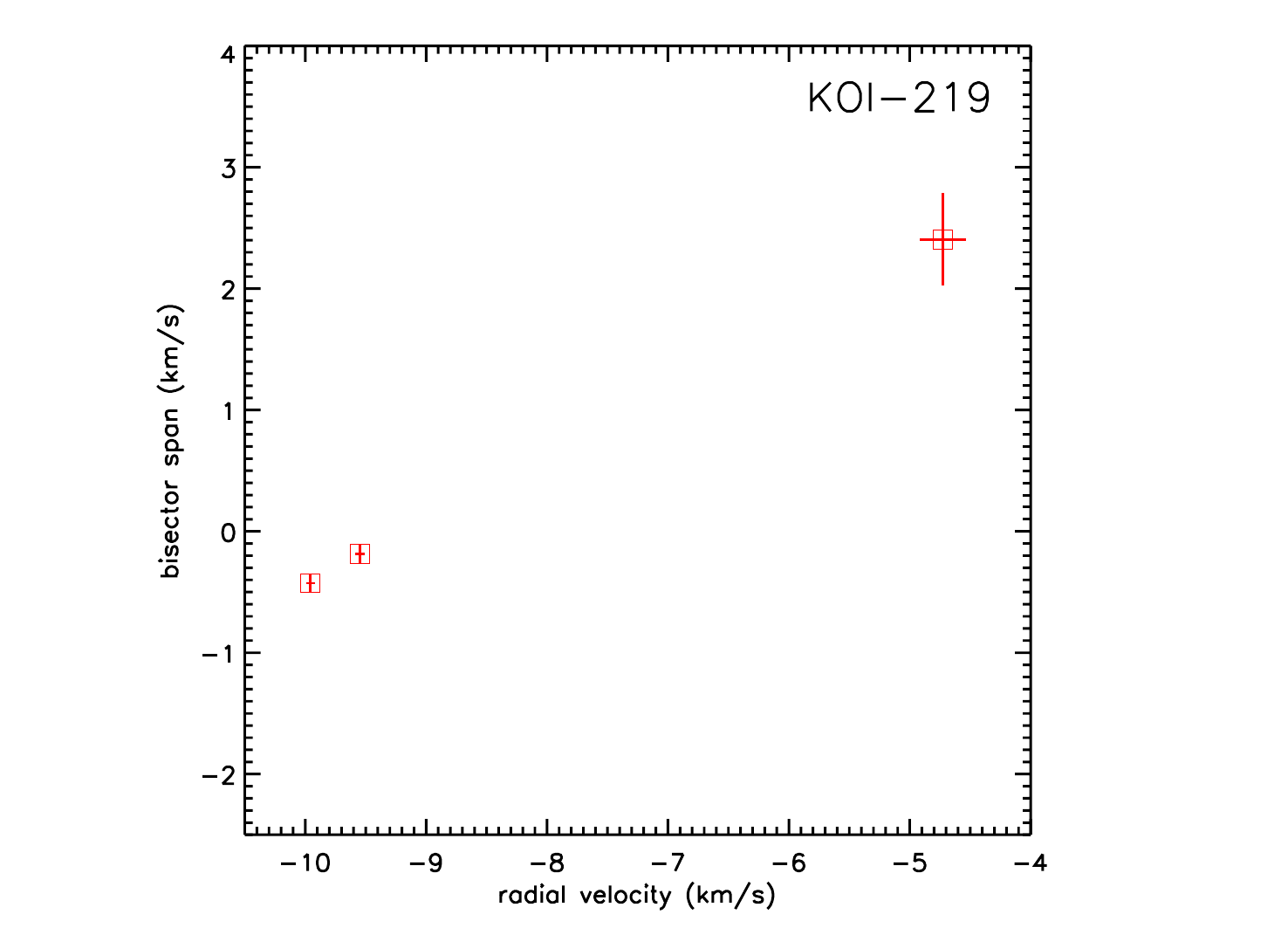}
\hspace{-2.19cm}
\includegraphics[scale=0.5]{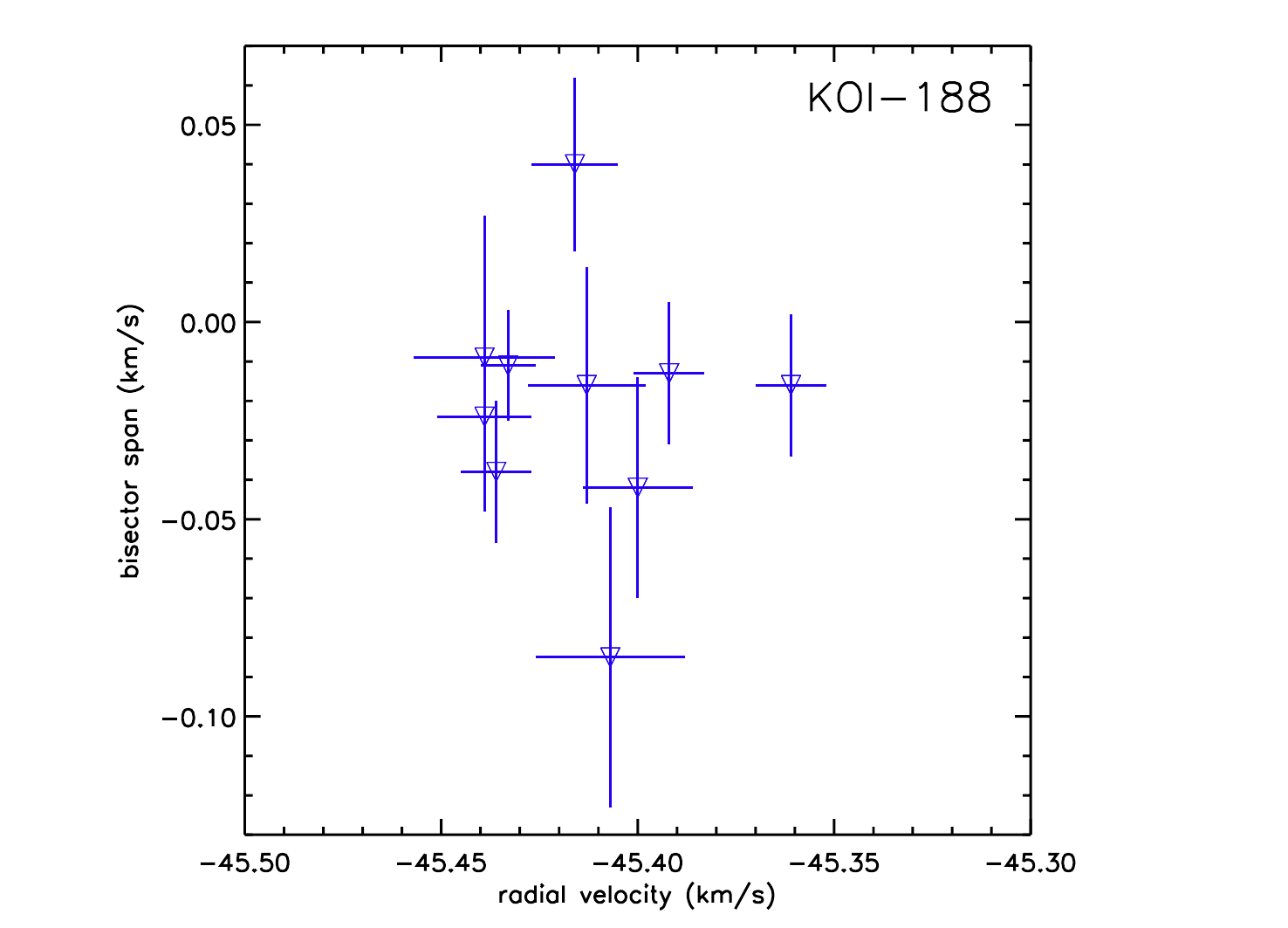}
\hspace{-2.19cm}
\includegraphics[scale=0.5]{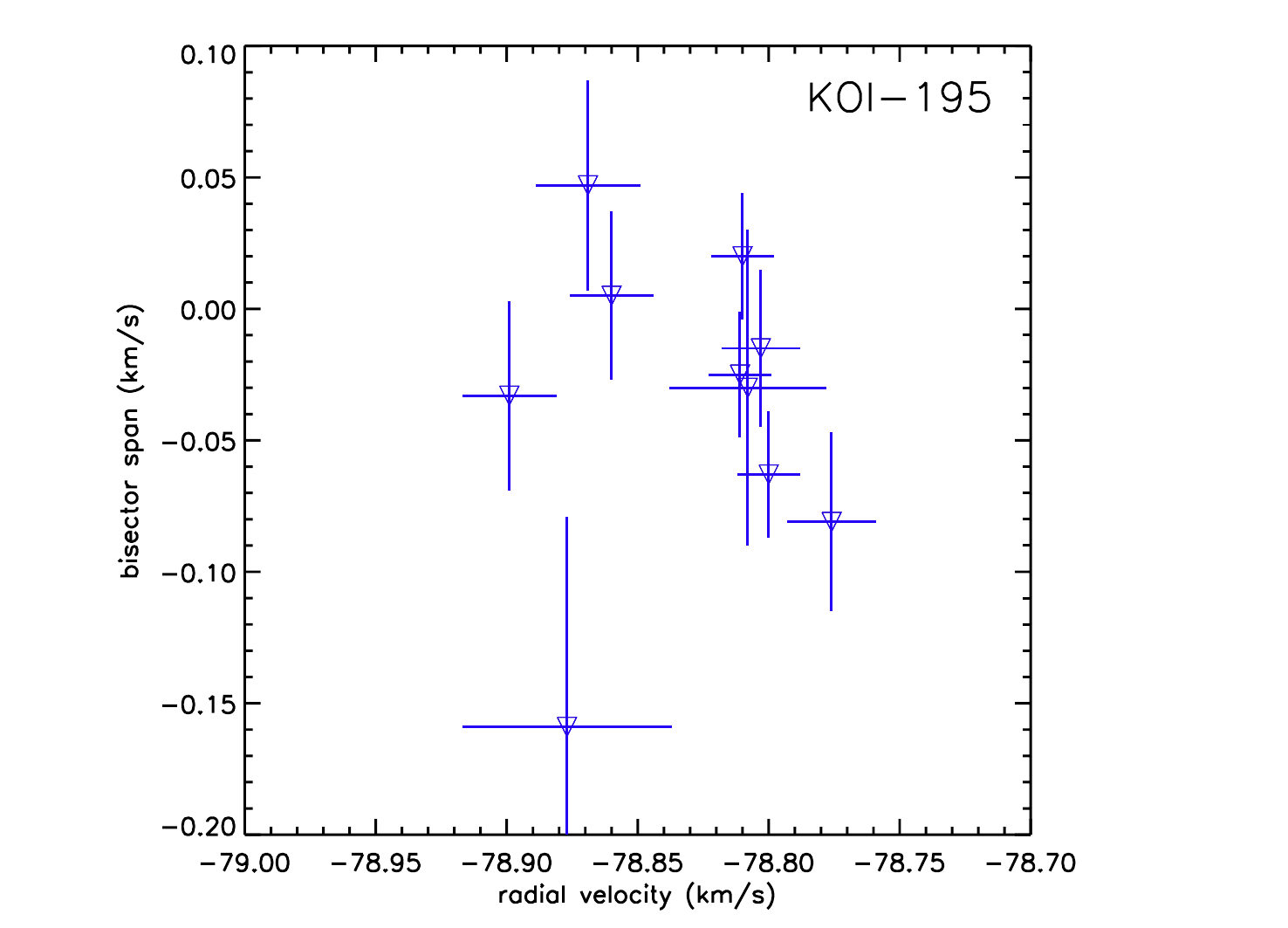}
\includegraphics[scale=0.5]{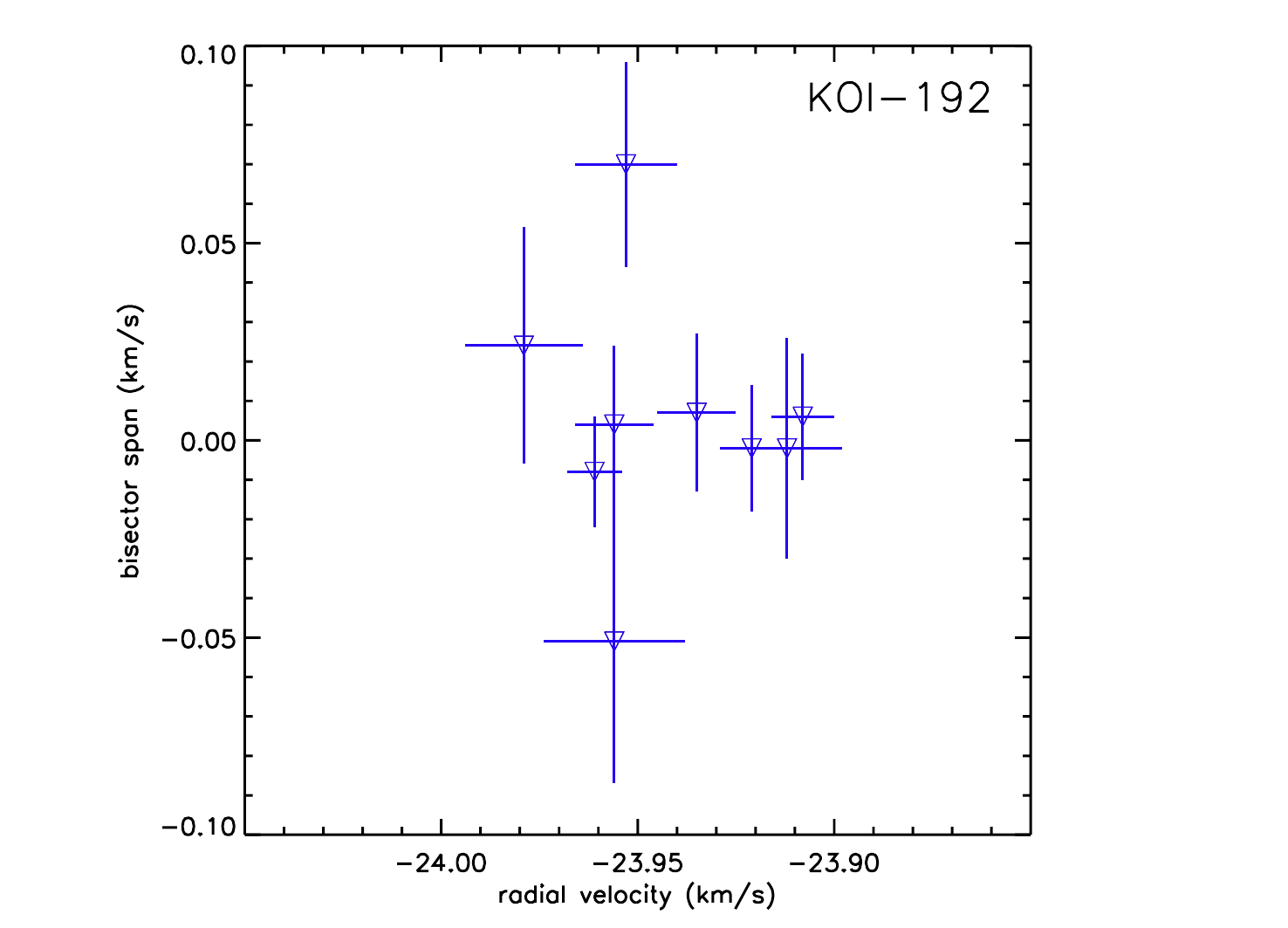}
\hspace{-2.19cm}
\includegraphics[scale=0.5]{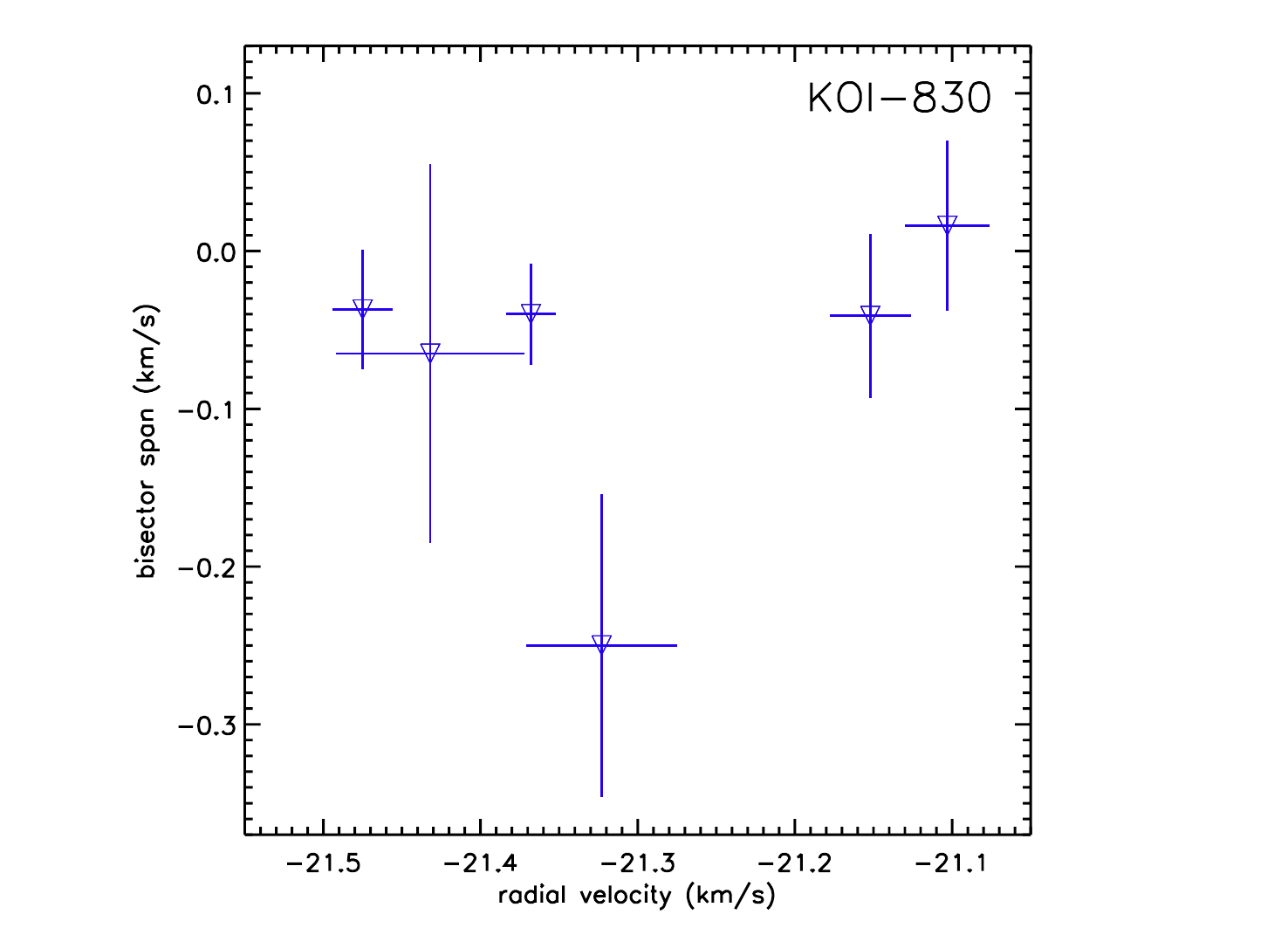}
  \caption{Bisector span as a function of the radial velocities with 1-$\sigma$\,Êerror bars
for the five targets. The ranges have the same extents in the $x$- and $y$-axes on each figure.
A correlation between bisector spans and radial velocities is clearly detected with SOPHIE on KOI-219, 
showing that the transiting planet candidate is a false positive. 
For the other four  targets, the HARPS-N measurements show no significant variations 
or trends of the bisector spans.
    }
  \label{fig_biss}
\end{figure*}

The SOPHIE observations of the three remaining candidates did not reveal any 
false positives. Two successive observations were performed 
in July 2011 for KOI-192, then in August 2012 for KOI-188 and KOI-195. 
For the three targets, the two observations 
were separated by a few days and made at quadratures, i.e., at $\sim P/4$ before and after a transit, 
$P$ being the period of the transits. This allows any radial velocity shift to be highest if the 
orbit of the transiting object is circular, which is a reasonable assumption given the short 
periods considered here. The SOPHIE radial velocities are plotted in the three upper panels of
Fig.~\ref{bestmodel_RV} and show no significant variations. The largest variation is observed in the 
case of KOI-188, but it is not significant according to the error~bars.

\begin{table}[t] 
\caption{SOPHIE pre-screening measurements.}
\begin{tabular}{cccrr}
\hline
BJD$_{\rm UTC}$ & RV & $\pm$$1\,\sigma$ & Texp$^\star$ & S/N$^\star$$^\star$ \\
-2\,456\,000 & [km/s] & [km/s]  & [sec] \\
\hline
\multicolumn{3}{l}{\hspace{-0.2cm}\emph{KOI-219:}}  \\
472.4354	&  --  	&	--		&	1800	&	10.8	\\
475.5494	&  --  	&	--		&	1800	&	14.3	\\
478.4095	&  -- 		&	--		&	1800	&	14.0	\\
\hline
\multicolumn{3}{l}{\hspace{-0.2cm}\emph{KOI-188:}}  \\
150.4740	&  -45.514  &	0.036	&	2115	&	13.5	\\
152.5412	&  -45.589  &	0.050	&	2513	&	11.6	\\
\hline
\multicolumn{3}{l}{\hspace{-0.2cm}\emph{KOI-195:}}  \\
153.4868	&  -78.976  &	0.046	&	1277	&	10.3	\\
154.5257	&  -78.991  &	0.043	&	1946	&	13.8	\\
\hline
\multicolumn{3}{l}{\hspace{-0.2cm}\emph{KOI-192:}}  \\
754.4488	&  -24.346  &	0.024	&	2274	&	17.3	\\
770.5076	&  -24.328  &	0.017	&	3304	&	17.4	\\
\hline
\multicolumn{5}{l}{$\star$: duration of each individual exposure.} \\
\multicolumn{5}{l}{$\star\star$: signal-to-noise ratio per pixel at 550\,nm.} \\
\label{table_rv_sophie}
\end{tabular}
\end{table}

\begin{table}[t] 
\caption{HARPS-N measurements.}
\begin{tabular}{cccrrr}
\hline
BJD$_{\rm UTC}$ & RV$^\dagger$ & $\pm$$1\,\sigma$ & bisect.$^\ddagger$ & Texp$^\star$ & S/N$^\star$$^\star$ \\
-2\,456\,000 & [km/s] & [km/s] & [km/s]  & [sec] \\
\hline
\multicolumn{3}{l}{\hspace{-0.2cm}\emph{KOI-188:}}  \\
155.4892  		&	-45.436		&	0.009	&	-0.038	&  2700	&	9.0    \\
157.3792  		&	-45.392		&	0.009	&	-0.013	&  2700	&	10.3  \\
185.5019  		&	-45.413		&	0.015	&	-0.016	&  2300	&	6.6   \\
186.3782  		&	-45.416		&	0.011	&	 0.040	&  2300	&	7.7   \\
489.5937  		&	-45.439		&	0.012	&	-0.024	&  2300	&	9.2   \\
490.6074  		&	-45.439		&	0.018	&	-0.009	&  2100	&	6.5   \\
491.6068  		&	-45.361		&	0.009	&	-0.016	&  2700	&	11.4  \\
492.6194  		&	-45.400		&	0.014	&	-0.042	&  2700	&	8.1   \\
493.6043  		&	-45.433		&	0.007	&	-0.011	&  2700	&	13.7  \\
494.6163  		&	-45.407		&	0.019	&	-0.085	&  2700	&	6.5   \\
\hline
\multicolumn{3}{l}{\hspace{-0.2cm}\emph{KOI-195:}}  \\
156.5564  		&	-78.800		&	0.012	&	-0.063	&  2700	&	10.2 \\
157.4515  		&	-78.860		&	0.016	&	 0.005	&  2700	&	7.8  \\
185.4254  		&	-78.776		&	0.017	&	-0.081	&  2700	&	6.9  \\
187.3785  		&	-78.877		&	0.040	&	-0.159	&  2300	&	3.4  \\
489.5329  		&	-78.899		&	0.018	&	-0.033	&  2300	&	8.8  \\
490.6366  		&	-78.808		&	0.030	&	-0.030	&  2300	&	5.7  \\
491.5386  		&	-78.810		&	0.012	&	 0.020	&  2700	&	11.9 \\
492.6521  		&	-78.869		&	0.020	&	 0.047	&  2700	&	7.7  \\
493.5702  		&	-78.811		&	0.012	&	-0.025	&  2700	&	12.4 \\
494.5483  		&	-78.803		&	0.015	&	-0.015	&  2700	&	10.0  \\
\hline
\multicolumn{3}{l}{\hspace{-0.2cm}\emph{KOI-192:}}  \\
156.3892  		&	-23.961		&	0.007	&	-0.008	&  2700	&	16.9 \\
159.4238  		&	-23.935		&	0.010	&	 0.007	&  2700	&	11.6 \\
185.5580  		&	-23.953		&	0.013	&	 0.070	&  1400	&	9.5  \\
188.3849  		&	-23.956		&	0.018	&	-0.051	&  2700	&	6.4  \\
490.5791  		&	-23.912		&	0.014	&	-0.002	&  2300	&	12.0 \\
491.5727  		&	-23.908		&	0.008	&	 0.006	&  2700	&	18.5 \\
492.5558  		&	-23.921		&	0.008	&	-0.002	&  2700	&	17.7 \\
493.6969  		&	-23.956		&	0.010	&	 0.004	&  2500	&	15.3 \\
494.5825  		&	-23.979		&	0.015	&	 0.024	&  2700	&	11.4 \\
\hline
\multicolumn{3}{l}{\hspace{-0.2cm}\emph{KOI-830:}}  \\
157.4875  		&	-21.368		&	0.016	&	-0.040	&  2700	&	5.8 \\
489.6748 		&	-21.152		&	0.026	&	-0.041	&  2100	&	4.6 \\
491.6366 		&	-21.475		&	0.019	&	-0.037	&  2100	&	6.2 \\
492.6819 		&	-21.323		&	0.048	&	-0.250	&  1800	&	2.5 \\
493.7245 		&	-21.103		&	0.027	&	 0.016	&  1800	&	4.5 \\
494.7325 		&	-21.432		&	0.060	&	-0.065	&  1300	&	2.4 \\
\hline
\multicolumn{6}{l}{$\dagger$:  radial velocities include season offsets 
$\delta V_0$ (see Sect.~\ref{sect_RV_harps}).}\\
\multicolumn{6}{l}{$\ddagger$: bisector spans; error bars are twice those of the RVs.} \\ 
\multicolumn{6}{l}{$\star$: duration of each individual exposure.} \\
\multicolumn{6}{l}{$\star\star$: signal-to-noise ratio per pixel at 550\,nm.} \\
\label{table_rv}
\end{tabular}
\end{table}

\subsubsection{Accurate radial velocities with HARPS-N}
\label{sect_RV_harps}

Since these three targets harbor no evident massive transiting objects, we observed them 
with HARPS-N in order to increase the radial velocity accuracy by comparison to SOPHIE.
We directly observed KOI-830 with HARPS-N, without initial pre-screening selection with SOPHIE.
The  HARPS-N spectrograph (Cosentino et al.~\cite{cosentino12}) is similar to SOPHIE, but
it allows better radial velocity accuracy to be reached, because of a more efficient 
stabilization system, and in particular in the case of our faint targets because of the larger size 
of its telescope, the 3.58~m TNG.
Its spectra~provide a larger resolution power $\lambda/\Delta\lambda=115\,000$. 
Our HARPS-N observations were secured in three different runs: five half-nights in 
August 2012, then four half-nights in September 2012, and finally six half-nights in 
July 2013. The three runs were made with the same instrument setup, with one 
exception: in 2012 we used the slow read-out mode of the detector, whereas 
in 2013 we used the fast read-out mode.
The two 1''-wide optical-fiber apertures were used, the first one being on the target 
whereas the second one was on the nearby sky to estimate the moonlight pollution. 
The second aperture shows detectable moonlight only on some~exposures secured 
at the end of the last three  half-nights of the 2013 run. Nevertheless, the apparent 
radial velocity of the Moon each time was far from that of the targets and weak enough to avoid~a 
significant effect on the radial velocity measurement.

As in the case of SOPHIE above, the HARPS-N spectra were extracted from the detector images 
with the standard DRS pipeline, which includes localization of the spectral 
orders on the 2D images, optimal order extraction, cosmic-ray rejection, corrections of flat-field, 
and wavelength calibration with thorium lamp exposures made during the afternoon. 
Then the spectra passed through weighted cross-correlation with 
numerical masks following the method described by 
Baranne et al.~(\cite{baranne96}) and Pepe et al.~(\cite{pepe02}). 
Because the blue part of the spectra has particularly low signal-to-noise ratios (S/N), 
we did not use the first 15 blue of the 70 available orders for the cross-correlation.
All the exposures provide a well-defined, single peak in the cross-correlation function (CCF), 
whose Gaussian fits allow the radial velocities to be measured together with their 
associated~uncertainties. The bisector spans of each CCF were also~measured. 

The HARPS-N measurements are reported in Table~\ref{table_rv} and are
plotted in Fig.~\ref{bestmodel_RV}. As seen on the phase-folded curves, 
for each of the four targets the radial velocity variations are in phase with 
the \kepler\ ephemeris and have semi-amplitudes of $\sim30$, 
 $\sim50$, $\sim30$, and $\sim200$~m/s for KOI-188, KOI-195, 
 KOI-192, and KOI-830, respectively. This corresponds to Saturn-mass 
companions for the first three  objects, and to a Jupiter-mass companion for 
the fourth one.

To fit the data we have to use an offset (or a drift) in addition to the Keplerian model adjusted to the \kepler\ ephemeris.
The offset being similar for the four targets (on the order of $-20$\,m/s/yr), this clearly indicates 
its origin is instrumental and not the signature of additional, long period companions in the systems. 
Exactly the same spectral orders were used in the cross-correlation, 
which in addition was done using the same version of the DRS pipeline (HARPN\_3.6), so there are 
no obvious reasons due to the reduction for such systematic shift. 
We found no systematic differences in the characteristics of the 2012 and 2013 measurements,
e.g., concerning their typical accuracies or the moonlight pollution.
The HARPS-N detector 
was changed between our 2012 and 2013 runs, but no shifts as large as the one we detect here
were seen in other targets observed both before and after that modification
(F.~Pepe, private communication). 
Another possibility is the charge transfer inefficiency (CTI) of the detector, which could 
induce radial velocity shifts at low S/N (e.g., Bouchy et al.~\cite{bouchy08}). 
This effect is not corrected on our data and is likely to be different between the 2012 
and 2013 HARPS-N detectors. 
However, our observations of a given target present various, low S/N, without correlations
between the S/N and the radial velocity residuals after the Keplerian fits. So the CTI 
effect seems to be low, for both  the 2012 and 2013 detectors.
The only difference we identify between our 
datasets is the read-out modes of the detector, 
which was the slower of the two 2012 runs, and the faster in 2013.
If this really  is the cause of the instrumental 
shift, this could be tested in the future by additional, dedicated tests on constant stars. 
So in the rest of the paper we assume there is an instrumental shift $\delta V_0$ 
between our 2012 and 2013 radial velocity measurements, 
which is computed  in Sect.~\ref{sect_stell_analys}.
The data  in Table~\ref{table_rv} and Figs.~\ref{bestmodel_RV} and~\ref{fig_biss}
are corrected for that shift. As the fast read-out mode is the standard one, we used 
the 2013 data as the reference ones. The data secured in slow read-out mode are 
redshifted in comparison to those secured in fast read-out mode,
so we corrected by shifting the first ones toward the blue.

The cross-correlation function bisector spans show neither significant variations nor trends 
as a function of radial velocity (Fig.~\ref{fig_biss}). This is the same for the width of the CCFs.
This agrees with the interpretation of the 
measured radial velocity variations as only being due to Doppler shifts instead of line profile 
deformations. Similarly, the radial velocities obtained using different stellar masks 
(G2, K0, or~K5) produce variations with similar amplitudes. So there is no evidence that 
the variations could be explained by  scenarios implying blended stars. We have chosen 
the numerical mask that produced the best fits, i.e., the less dispersed residuals around the 
Keplerian fits: there are the K5-, G2-, K0-, and K0-type masks for KOI-188, KOI-195, KOI-192, 
and KOI-830, respectively. 

All these observations allow us to conclude that the four~targets harbor transiting giant planets,
and so
hereafter we  do~not~consider them anymore as candidates but as planets, which we~now
designate as KOI-188b, KOI-195b, KOI-192b, and~KOI-830b.

\section{System characterization}
\label{sect_analysis}

\subsection{Spectral analysis of the host stars}
\label{sect_stell_analys}

We analyzed the co-added individual HARPS-N spectra of the four planet-host stars.
The signal-to-noise ratios of the co-added spectra 
are 160, 120, 220, and 60 per resolution 
element~at 550~nm in the continuum for KOI-188, KOI-195, KOI-192, and KOI-830, respectively.
The projected rotational velocity \vsini\ was determined from a set of isolated spectral~lines.
None of the stars is rapidly rotating.

We performed the spectral analysis
using the iterative spectral synthesis package VWA (versatile wavelength analysis). 
As described in 
detail by Bruntt et al.~(\cite{bruntt10}) and references therein, 
the atmospheric parameters \teff, \logg, and \met\ were derived from tens of 
\ion{Fe}{i} and \ion{Fe}{ii} weak lines carefully selected, 
the exact number of lines depending on the signal-to-noise ratio of the spectra.
Ionization and excitation equilibria were imposed, as well as a zero slope between 
the abundances given by individual lines and their equivalent widths. 
As a verification, we also derived the surface gravity 
from the \ion{Ca}{i} pressure-sensitive line at 612.2\,nm. 
These parameters are used below  (Sect.~\ref{sect_parameter_system}) 
to derive the fundamental parameters of the four stars (mass, radius, and age)
from the distribution of stellar density derived from the transit modeling  
and from the comparison of the location of the star in the H-R Diagram with
evolution tracks.
As the S/N of the co-added HARPS-N spectrum of KOI-830 is lower than 
that of the other objects, this implies less accurate derived stellar parameters 
for that star. This is particularly the case for \vsini\  and  \logg\ which are difficult 
to measure with the available data, and for which we chose to adopt conservative 
error bars.

\begin{table*}[th]
\centering
\caption{Parameters of the transiting hot Saturns KOI-188b and KOI-195b and their host stars.}            
\hspace{-1.5cm}
\begin{minipage}{17cm} 
\setlength{\tabcolsep}{1cm}
\renewcommand{\footnoterule}{}                          
\begin{tabular}{lcc}        
\hline                
 							& KOI-188 				& KOI-195			\\
\hline
\multicolumn{3}{l}{\hspace{-0.8cm}\emph{Ephemerides and orbital parameters:}} \\
Planet orbital period $P$ [days] 			& $3.79701816 \pm 0.00000019$  		& $3.21751883 \pm 0.00000019$	\\ 
Transit epoch $T_{0}$ [BJD -- 2\,454\,900] 		& $66.508785 \pm 0.000039$ 		& $66.631964 \pm 0.000047$ 	\\  
Orbital eccentricity $e$ (99\,\%\ upper limit)	& $<0.33$				& $<0.18$			\\
Orbital inclination $i_p$ [$^{\circ}$] 			& $87.02 \pm 0.08$ 			& $85.74 \pm 0.06$		\\  
Transit duration $T_{1-4}$ [hours] 			& $2.353 \pm 0.006$ 			& $2.191 \pm 0.010$ 		\\
Impact parameter $b$ 					& $0.602 \pm 0.012$  			& $0.718 \pm 0.007$  		\\
\hline
\multicolumn{3}{l}{\hspace{-0.8cm}\emph{Transit-related parameters:}} \\
System scale $a/R_{\star}$ 				& $11.60 \pm 0.08$ 			& $9.67 \pm 0.06$ 		\\
Radius ratio $k=R_{\rm p}/R_{\star}$ 			& $0.1168 \pm 0.0008$ 	 		& $0.1221 \pm 0.0013$ 		\\
Linear limb-darkening coefficient u$_{a}$ 		& $0.53 \pm 0.07$ 			& $0.43 \pm 0.20$ 		\\
Quadratic limb-darkening coefficient u$_{b}$ 		& $0.15 \pm 0.12$ 			& $0.20 \pm 0.27$ 		\\
\hline
\multicolumn{3}{l}{\hspace{-0.8cm}\emph{RV-related parameters:}} \\
Semi-amplitude $K$ [\ms] 				& $34 \pm 10$ 				& $50 \pm 10$ 			\\
HARPS-N systemic radial velocity  $V_{0}$ [\kms] 	& $-45.404 \pm 0.009$ 			& $-78.835 \pm 0.010$		\\
HARPS-N season offset  $\delta V_0$ [\ms] 		& $29 \pm 18$ 				& $22 \pm 20$ 			\\
O-C residuals [\ms] 					& 11.5 					& 8.6				\\
\hline
\multicolumn{3}{l}{\hspace{-0.8cm}\emph{Data-related parameters:}} \\
\kepler\ season 0 contamination [\%] 			& $4.50 \pm 0.40$			& $5.78  \pm 0.42$ 		\\
\kepler\ season 1 contamination [\%] 			& $4.30 \pmÊ0.43$ 			& $5.78  \pmÊ0.43$		\\
\kepler\ season 2 contamination [\%] 			& $4.42Ê\pm 0.41$ 			& $6.16  \pm 0.45$		\\
\kepler\ season 3 contamination [\%] 			& $4.28Ê\pmÊ0.40$ 			& $6.48  \pmÊ0.44$		\\
\kepler\ season 0 jitter LC [ppm] 			& $121 \pm 11$ 				& $204Ê\pm 10$			\\
\kepler\ season 0 jitter SC [ppm] 			& $310 \pm 46$ 				& $66Ê\pm 59$			\\
\kepler\ season 1 jitter LC [ppm] 			& $99 \pm 12$ 				& $246 \pm 9$			\\
\kepler\ season 1 jitter SC [ppm] 			& $156 \pm 85$ 				& $-$				\\
\kepler\ season 2 jitter LC [ppm] 			& $240 \pm 9$ 				& $315 \pmÊ11$			\\
\kepler\ season 2 jitter SC [ppm] 			& $36  \pm 33$ 				& $210 \pmÊ9$			\\
\kepler\ season 3 jitter LC [ppm] 			& $164 \pm 10$ 				& $237 \pm 15$			\\
\kepler\ season 3 jitter SC [ppm] 			& $46  \pm 41$ 				& $98 \pm 69$			\\
HARPS-N jitter [\ms] 					& $0.014 \pm 0.015$			& $0.011 \pm 0.015$		\\ 
SED jitter [mag] 					& $0.065 \pm 0.026$ 			& $0.084 \pm 0.029$ 		\\ 
\hline
\multicolumn{3}{l}{\hspace{-0.8cm}\emph{Spectroscopic parameters:}} \\
Effective temperature \teff [K]				& $ 5170   \pm 70   $			& $ 5725   \pm 90   $   	\\
Metallicity \met\ [dex] 				& $ 0.24   \pm 0.11   $ 		& $ -0.21   \pm 0.08   $   	\\   
Stellar rotational velocity {\vsini} [\kms] 		& $3 \pm 1$ 				& $3 \pm 1$ 			\\   
Spectral type 						& K1V 					& G1V 				\\
Stellar surface gravity \logg\ [$g\;cm^{-2}$]		& $ 4.50   \pm 0.15  $		& $ 4.50   \pm 0.15   $	\\   
\hline
\multicolumn{3}{l}{\hspace{-0.8cm}\emph{Stellar physical parameters from combined analysis:}} \\
Stellar surface gravity \logg\ [$g\;cm^{-2}$]		& $ 4.535   \pm 0.010  $		& $ 4.471   \pm 0.011   $	\\   
Stellar density $\rho_{\star}$ [$g\;cm^{-3}$] 		& $1.448 \pm 0.030$ 			& $1.170 \pm 0.023$		\\  
Star mass $M_\star$ [\Msun] 				& $0.93 \pm 0.05$			& $0.91 \pm 0.06$		\\     
Star radius $R_\star$ [\Rsun] 				& $ 0.86\pm 0.02$   			& $ 0.92\pm 0.02$		\\  
Age of the star [Gyr] 					& $5 \pm 4$  	 			& $6 \pm 4$			\\     
Luminosity of the star $\log (L/L_{\odot})$      	& $-0.323 \pm 0.031$			& $-0.089 \pm 0.042 $		\\
Distance of the system [pc] 				& $650 \pm 20$ 				& $880 \pm 30$  		\\  
Interstellar extinction $E(B-V)$ 	[mag]		& $0.024 \pm 0.022$ 			& $0.055 \pm 0.030$		\\  
\hline
\multicolumn{3}{l}{\hspace{-0.8cm}\emph{Planetary physical parameters from combined analysis:}} \\
Orbital semi-major axis $a$ [AU] 			& $0.0464 \pm 0.0008$ 			& $0.0414 \pm 0.0010$ 		\\ 
Planet mass $M_{\rm p}$ [\Mjup] 			& $0.25 \pm 0.08$			& $0.34 \pm 0.08$ 		\\ 
Planet radius $R_{\rm p}$ [\Rjup]  			& $0.978 \pm 0.022$			& $1.09 \pm 0.03$ 		\\
Planet density $\rho_{\rm p}$ [g\,cm$^{-3}$] 		& $0.27 \pm 0.07$ 			& $0.26 \pm 0.06$ 		\\ 
Planetary equilibrium temperature  $T_{\mathrm{p}}$ [K]	& $1070 \pm 15  $			& $1300 \pm 20   $   		\\              
\hline
\vspace{-0.5cm}
\end{tabular}
\end{minipage}
\label{posterior_1}   
\vspace{-0.075cm}
\end{table*}

\begin{table*}[th]
\centering
\caption{Parameters of the transiting planets KOI-192b and KOI-830b and their host stars.}            
\hspace{-1.5cm}
\begin{minipage}{17cm} 
\setlength{\tabcolsep}{1cm}
\renewcommand{\footnoterule}{}                          
\begin{tabular}{lcc}        
\hline                
 							& KOI-192 				& KOI-830			\\
\hline
\multicolumn{3}{l}{\hspace{-0.8cm}\emph{Ephemerides and orbital parameters:}} \\
Planet orbital period $P$ [days] 			& $10.2909940 \pm 0.0000011$ 		& $3.52563254 \pm 0.00000015$	\\ 
Transit epoch $T_{0}$ [BJD -- 2\,454\,900] 		& $70.02207 \pm 0.00009$ 		& $103.048008 \pm 0.000032$ 	\\  
Orbital eccentricity $e$ (99\,\%\ upper limit)	& $<0.57$				& $<0.22$			\\
Orbital inclination $i_p$ [$^{\circ}$] 			& $89.50 \pm 0.45$ 			& $89.36 \pm 0.43$		\\  
Transit duration $T_{1-4}$ [hours] 			& $4.286 \pm 0.009$ 			& $2.614 \pm 0.007$		\\
Impact parameter $b$					& $0.09 \pm 0.07$ 			& $0.13 \pm 0.09$		\\
\hline
\multicolumn{3}{l}{\hspace{-0.8cm}\emph{Transit-related parameters:}} \\
System scale $a/R_{\star}$ 				& $14.2 \pm 2.1$ 			& $11.65 \pm 0.10$		\\
Radius ratio $k=R_{\rm p}/R_{\star}$ 			& $0.0913 \pm 0.0003$			& $0.1384 \pm 0.0010$		\\
Linear limb-darkening coefficient u$_{a}$ 		& $0.450\pm0.017$ 			& $0.584 \pm 0.017$ 		\\
Quadratic limb-darkening coefficient u$_{b}$ 		& $0.13 \pm 0.04$ 			& $0.10 \pm 0.06$		\\
\hline
\multicolumn{3}{l}{\hspace{-0.8cm}\emph{RV-related parameters:}} \\
Semi-amplitude $K$ [\ms] 				& $29.8 \pm 9.1$	 		& $188 \pm 26$ 			\\
HARPS-N systemic radial velocity  $V_{0}$ [\kms] 	& $-23.938 \pm 0.010$	 		& $-21.302 \pm 0.024$ 		\\
HARPS-N season offset  $\delta V_0$ [\ms] 		& $20 \pm 18$ 				& $3 \pm 31$ 			\\
O-C residuals [\ms] 					& 8.0 					& 13.4 				\\
\hline
\multicolumn{3}{l}{\hspace{-0.8cm}\emph{Data-related parameters:}} \\
\kepler\ season 0 contamination [\%] 			& $3.32 \pm 0.44$ 			& $7.68 \pm 0.52$ 		\\
\kepler\ season 1 contamination [\%] 			& $3.61 \pmÊ0.42$ 			& $7.24 \pmÊ0.52$		\\
\kepler\ season 2 contamination [\%] 			& $3.16Ê\pm 0.43$ 			& $7.07 \pm 0.52$		\\
\kepler\ season 3 contamination [\%] 			& $3.67Ê\pmÊ0.44$ 			& $7.46 \pmÊ0.51$		\\
\kepler\ season 0 jitter LC [ppm] 			& $150 \pm 8$ 				& $226Ê\pm 10$			\\
\kepler\ season 0 jitter SC [ppm] 			& $230 \pm 50$ 				& $-$				\\
\kepler\ season 1 jitter LC [ppm] 			& $127 \pm 9$ 				& $153 \pm 13$			\\
\kepler\ season 1 jitter SC [ppm] 			& $241 \pm 65$ 				& $-$				\\
\kepler\ season 2 jitter LC [ppm] 			& $101 \pm 10$ 				& $182 \pmÊ12$			\\
\kepler\ season 2 jitter SC [ppm] 			& $291 \pm 35$ 				& $-$				\\
\kepler\ season 3 jitter LC [ppm] 			& $122 \pm 9$ 				& $127 \pm 17$			\\
\kepler\ season 3 jitter SC [ppm] 			& $345 \pm 28$ 				& $-$				\\
HARPS-N jitter [\ms] 					& $0.008 \pm 0.012$			& $0.026 \pm 0.024$		\\ 
SED jitter 	 [mag]					& $0.056 \pm 0.025$ 			& $0.065 \pm 0.029$		\\ 
\hline
\multicolumn{3}{l}{\hspace{-0.8cm}\emph{Spectroscopic parameters:}} \\
Effective temperature \teff[K] 				& $ 5800  \pm 70  $			& $ 5150  \pm 100  $		\\
Metallicity \met\ [dex] 				& $ -0.19 \pm 0.07   $			& $ 0.09 \pm 0.17   $ 		\\   
Stellar rotational velocity {\vsini} [\kms] 		& $3 \pm 1$ 				& $2 \pm 2$ 			\\   
Spectral type 						& G2V 					& K1V 				\\
Stellar surface gravity \logg\ [$g\;cm^{-2}$]		& $ 4.15   \pm 0.15  $		& $ 5.0   \pm 0.4   $	\\   
\hline
\multicolumn{3}{l}{\hspace{-0.8cm}\emph{Stellar physical parameters from combined analysis:}} \\
Stellar surface gravity \logg\  [$g\;cm^{-2}$]		& $4.14 \pm 0.12$ 			& $4.571 \pm 0.011$ 		\\   
Stellar density $\rho_{\star}$ [$g\;cm^{-3}$] 		& $0.4 \pm 0.2$ 			& $1.704 \pm 0.041$ 		\\  
Star mass $M_\star$ [\Msun] 				& $0.96 \pm 0.06$			& $0.87 \pm 0.05$		\\     
Star radius $R_\star$ [\Rsun] 				& $1.35  \pm 0.20  $			& $0.80  \pm 0.02$		\\  
Age of the star [Gyr] 					& $7 \pm 4$				& $5\pm4$			\\     
Luminosity of the star $\log (L/L_{\odot})$      	& $0.29 \pm 0.13$			& $-0.39 \pm 0.05$		\\
Distance of the system [pc] 				& $1100 \pm 150$			& $720 \pm 25$ 			\\  
Interstellar extinction $E(B-V)$ 	[mag] 		& $0.022 \pm 0.019$ 			& $0.040 \pm 0.037$ 		\\  
\hline
\multicolumn{3}{l}{\hspace{-0.8cm}\emph{Planetary physical parameters from combined analysis:}} \\
Orbital semi-major axis $a$ [AU] 			& $0.091 \pm 0.010$			& $0.0433 \pm 0.0009$		\\ 
Planet mass $M_{\rm p}$ [\Mjup] 			& $0.29 \pm 0.09$  			& $1.27 \pm 0.19$ 		\\ 
Planet radius $R_{\rm p}$ [\Rjup]  			& $1.23 \pm 0.21$ 			& $1.08 \pm 0.03$		\\
Planet density $\rho_{\rm p}$ [g\,cm$^{-3}$] 		& $0.16 \pm 0.14$ 			& $1.02 \pm 0.15$ 		\\ 
Planetary equilibrium temperature  $T_{\mathrm{p}}$ [K]	& $1100 \pm 70  $			& $1070 \pm 25$			\\              
\hline
\vspace{-0.5cm}
\end{tabular}
\end{minipage}
\label{posterior_2}   
\vspace{-0.075cm}
\end{table*}

\subsection{Parameters of the planetary systems}
\label{sect_parameter_system}

The normalized \kepler\ light curves were fitted together with the HARPS-N radial velocities 
by a transit model using the EBOP code (Etzel~\cite{etzel81}) and a Keplerian model.
For the combined fits we used the \texttt{PASTIS} code (D\'{\i}az et al.~\cite{diaz14}), following 
the procedures used, e.g., by D\'{\i}az et al.~(\cite{diaz13}) 
and H\'ebrard et al.~(\cite{hebrard13a}). 
The analysis also includes the fit of the SED 
(from magnitudes reported in Table~\ref{startable_KOI}) 
and stellar evolution tracks to determine coherent stellar parameters.
The distances of the four stars were determined  by comparing the 
SED with an interpolated grid of synthetic spectra 
from the PHOENIX/BT-Settl library (Allard et al.~\cite{allard12}), corrected for the interstellar 
extinction.
We used four different stellar evolution tracks as input for the stellar parameters of each target: 
StarEvol (Lagarde et al.~\cite{lagarde12}; A.~Palacios, private communication), 
Dartmouth (Dotter et al.~\cite{dotter08}), 
Parsec (Bressan et al.~\cite{bressan12}), and
Geneva (Mowlavi et al.~\cite{mowlavi12}).

We used an oversampling factor 
of ten when comparing the model with the \kepler\  long-cadence light curves to account for their 
long integration time (Kipping~\cite{kipping10}; Kipping \& Bakos~\cite{kipping11}). 
As explained above (see Sect.~\ref{sect_RV_harps}), we allow a free radial velocity shift 
between the 2012 and 2013 HARPS-N data in order to correct for the 
season offset  $\delta V_0$.
For each \kepler\  light curve we 
included the out-of-transit flux and the contamination factor as free parameters. 
Contamination factors reported in the KIC (Brown et al.~\cite{brown11}) 
have been shown to be incorrect in some cases  
(see, e.g., KOI-205; D\'{\i}az et al.~\cite{diaz13}) 
so we chose to fit them instead of adopting the KIC values.
Similarly, we left the limb-darkening coefficients free to vary within a Gaussian prior 
centered on the estimated values.
We also account for additional sources of Gaussian noise in the light curves, radial velocities, 
and SED by fitting a jitter value to each dataset. This is  especially appropriate for the \kepler\ data 
since the star is located on different CCDs each season. 
For each target we performed both fits with eccentric or circular orbits. In none of them did we detect 
a significant eccentricity. In the case of KOI-192 we obtained a weak constraint on the eccentricity 
so we adopted the conservative values of the parameters obtained with the eccentric fit. For the 
three other targets we constrained the eccentricities to low values so we adopted the parameters 
obtained from the circular fits; we discuss that assumption below.
We conservatively adopted the 99\,\%\ upper limits for the eccentricity of the four~systems.

For each of the four systems and datasets we thus have 25 free parameters (27 for eccentric fits), 
which we have fitted using a Metropolis-Hasting Markov chain Monte Carlo (MCMC) 
algorithm (e.g., Tegmark et al.~\cite{tegmark04}) 
with an adaptive step size (Ford~\cite{ford06}). To better sample the posterior 
distribution in the case of non-linear correlations between parameters, we applied an 
adaptive principal component analysis to the chains and jumped the parameters in 
an uncorrelated space (D\'{\i}az et al.~\cite{diaz14}).
For most of the parameters of the MCMC we used non-informative priors (uniform or Jeffreys distributions).
Exceptions are 
the stellar parameters \teff, \met, and $\rho_{\star}$ derived from the above spectral analysis 
(Sect.~\ref{sect_stell_analys}), and 
the orbital periods and phases of the planets for which we used as priors the \kepler\
values with error bars increased by a factor of 100 to avoid biases. 
We chose the factor 100 as a conservative one by comparison with smaller factors which 
provide similar results. We finally obtained orbital periods in perfect agreement with the 
\kepler\ values.
All the priors of the final fits are listed in the Table~\ref{PriorTable} available online.

The systematic uncertainties due to the stellar evolution model are not easy to quantify. 
In order to attempt to take these uncertainties into account in our final results, the 
MCMC chains were calculated using the four different sets of models in equal proportions. 
This allows possible discrepancies between stellar evolution models to be taken into 
account in the MCMC chains and thus in our final results, 
while including the state-of-the-art knowledge about stellar evolution models.
Some additional uncertainties might remain, as it could be defaults common to all stellar 
evolution~models. 

Each system was analyzed with $4\times10$ chains (one per stellar evolution track)
leading to a total of $4\times10^{7}$ steps. 
Each chain was started at random points drawn from the joint prior. 
All chains converged to the same solution. 
For each target
we computed the correlation length of the converged sub-chains before thinning them. 
We finally merged the thinned chains, which left us with a total of more than 1000 independent 
samples of the posterior distribution for each~target. 

The models are shown in Figs.~\ref{bestmodel_LC} and~\ref{bestmodel_RV}, and 
the 68.3\,\%\ confidence intervals (corresponding to 1-$\sigma$ intervals assuming 
Gaussian distributions) are listed in Tables~\ref{posterior_1} and~\ref{posterior_2}.
We did not find any significant jitter to add to radial velocities, and 
the dispersion of the residuals around the Keplerian fit are slightly smaller than the 
error bars on the radial velocities. This suggests the estimated uncertainties on the 
HARPS-N radial velocities could be slightly overestimated. 
The four measured values of the instrumental offsets between 2012 and 2013 provide 
the average shift $\delta V_0 = 21.6 \pm 10.2$~m/s. The fact that data secured in 
slow read-out mode are redshifted by this offset value in comparison to those secured in 
fast read-out mode can be checked by dedicated~tests.
We did not use the SOPHIE radial velocities in the final fit because their error bars 
do not allow significant constraints to be put by comparison to the more accurate 
HARPS-N radial velocities. 
Still, the SOPHIE data are plotted in Fig.~\ref{bestmodel_RV} for illustration after their 
systemic radial velocity was determined from $\chi^2$ variations (we found 
$-45.551 \pm 0.035$, $-78.883 \pm 0.022$, and $-24.342 \pm 0.018$\,\kms\ 
for KOI-188, KOI-195, and KOI-192, respectively).

The out-of-transit fluxes were found around unity with typical uncertainties of 10\,ppm.
The contamination factors found for short- and long-cadence data agree for each 
season of each target. We found values similar to those tabulated in the KIC, except 
for KOI-195 where our fitted contamination factors were systematically larger than the 
KIC ones.
The jitters found on the photometry are classical for \kepler\ data. 
Our fitted limb-darkening coefficients are compatible within 1\,$\sigma$ with the expected 
values from Claret \&~Bloemen~(\cite{claret11}).
None of the light curves show the signature of planetary occultation at the secondary eclipse 
phase, as expected for relatively long-period planets such as these.

Finally, we investigated the bulk composition of the four planets using CEPAM 
(e.g.,~Guillot~\cite{guillot10}).
Planetary evolution models have been built following the method 
used in, e.g., Deleuil et al.~(\cite{deleuil14}). The planets are assumed to be made of a central rocky 
core surrounded by a solar-composition envelope, but because they are irradiated giant planets, 
we also considered the cases where we dissipate a fraction (1\%) of the incoming stellar 
flux deep in the layers of the planet (for more details see, e.g., Guillot~\&\ Havel~\cite{guillot11}; 
Almenara et al.~\cite{almenara13}).

\section{Results and discussions}
\label{sect_disc}

\begin{figure}[]
\begin{center}
\includegraphics[width=\columnwidth]{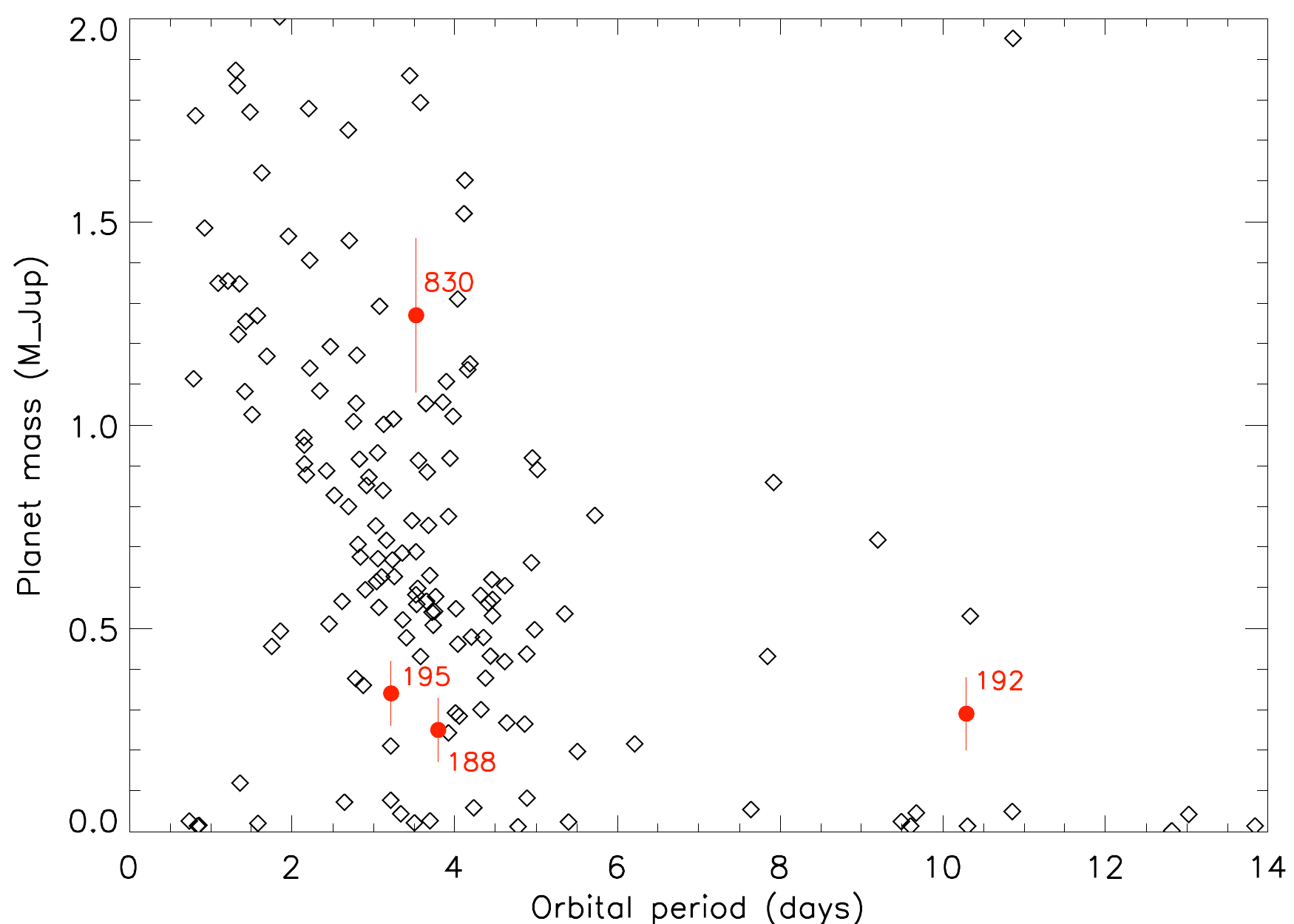} 
\includegraphics[width=\columnwidth]{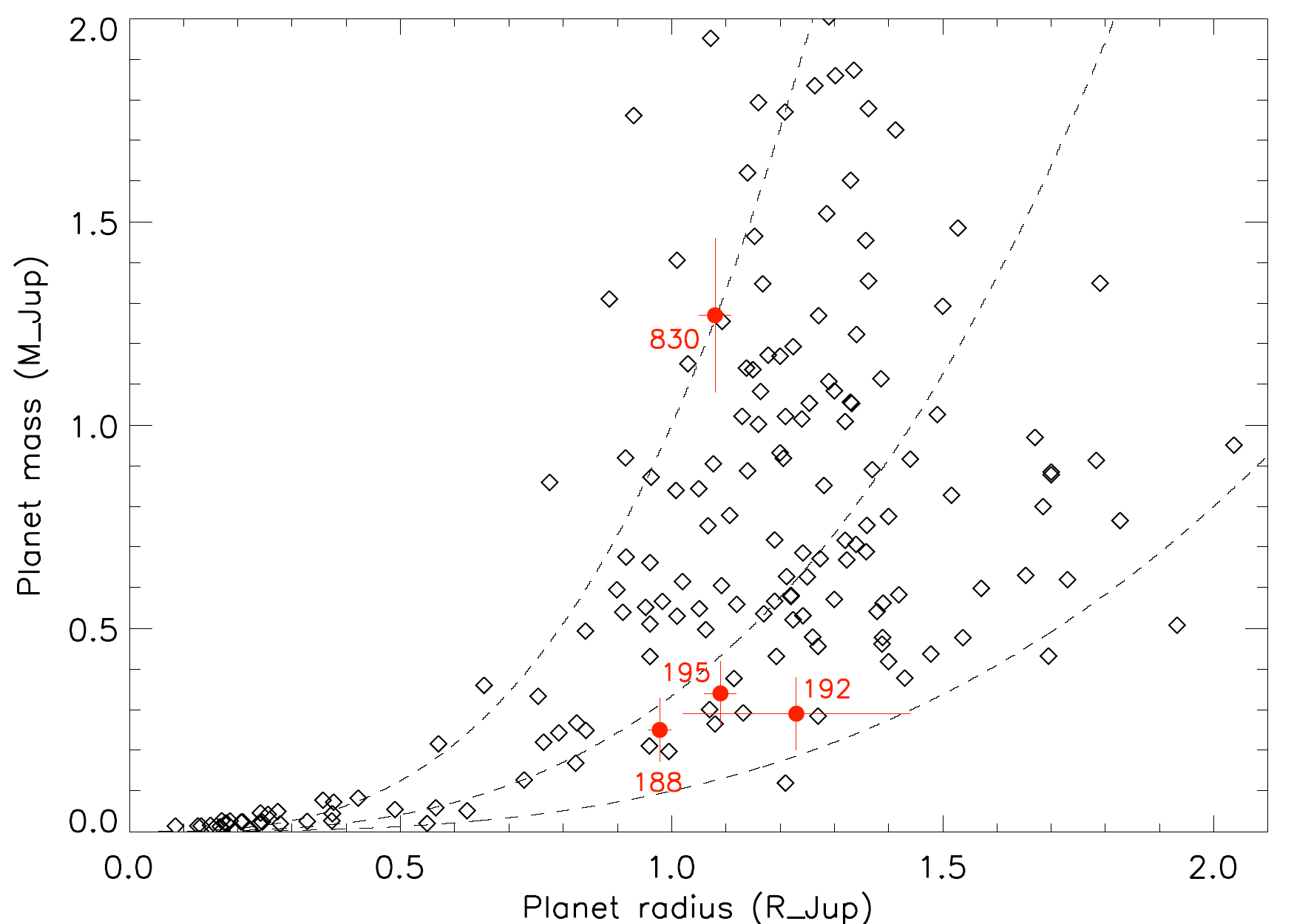} 
\caption{Masses and orbital periods (top) 
and masses and radii (bottom)
of the four new planets
KOI-188b, KOI-195b, KOI-192b, and KOI-830b (red circles), 
compared to other known transiting, close-in planets (black diamonds).
In the bottom panel, the three dashed curves show the Jupiter density, and
1/3 and 1/10 of that value (from left to right).
}
\label{fig_stat}
\end{center}
\end{figure}

The four new, transiting planets presented here are giant, close-in ones, with radii on the 
order of the Jupiter radius. They orbit slow-rotating stars
located on the main sequence of the H-R Diagram.
Their masses, periods, and radii are plotted in Fig.~\ref{fig_stat}, 
and compared 
with other transiting planets as known in May 2014 according to the Exoplanet Orbit Database 
(exoplanets.org; Wright et al.~\cite{wright11}).
Our discussion below concentrates on giant planets mainly shown in these plots, 
and not on the low-mass and small-radius planets seen in the lower parts of 
the figures, which are clearly different from the four new planets presented~here.

KOI-188b and KOI-195b are similar planets: their periods are $3.79701816 \pm 0.00000019$  and 
$3.21751883 \pm 0.00000019$~days, their masses $0.25 \pm 0.08$	 and $0.34 \pm 0.08$\,\Mjup, 
and their radii $0.978 \pm 0.022$ and $1.09 \pm 0.03$\,\Rjup, respectively. The upper limits we found at 99\,\%\ 
on their eccentricities are 0.33 and 0.18. Eccentric fits provide similar results to the 
circular ones presented above, but with slightly increased uncertainties for some parameters such as 
the inclination or the $a/R_{\star}$ ratio. These close-in planets are expected to be circularized and 
to have eccentricities near 0.
Indeed, even considering the higher bounds for their eccentricities, their circularization time-scales 
would be on the order of a few hundreds of Myrs, assuming a typical tidal dissipation efficiency of 
$10^6$ and $10^7$ for the planets and the stars, respectively (Matsumura et al.~\cite{matsumura08}). 
These two planets are similar to standard 
hot Jupiters but with lower masses, and could be called hot Saturns. Analogous planets include 
WASP-29b, WASP-39b, WASP-49b, HAT-P-19b, or CoRoT-25b
(Hellier et al.~\cite{hellier10}; Faedi et al.~\cite{faedi11}; Lendl et al.~\cite{lendl12}; 
Hartman et al.~\cite{hartman11}; Almenara et al.~\cite{almenara13}).
As seen in the upper panel of Fig.~\ref{fig_stat}, they are located in the low-mass range of the envelope 
where planets with shorter periods are rare. The lack of hot Saturns could be interpreted as a 
signature of evaporation of planets too close to their stars (e.g., Lecavelier des \'Etangs~\cite{lecavelier97};
Ehrenreich \&\ D\'esert~\cite{ehrenreich11b}). 
Both planets have particularly low densities on the order of $\rho_{\rm p} = 0.25$~g\,cm$^{-3}$
(lower panel of Fig.~\ref{fig_stat}), making them favorable 
targets for atmospheric species detection in~absorption. They could be considered as bloated 
hot Saturns as CoRoT-25b (Almenara et al.~\cite{almenara13}), 
by comparison with planets with similar periods and masses but smaller radii and larger densities, 
such as CoRoT-8b (Bord\'e~et al.~\cite{borde10}). 

Both planets are slightly inflated  compared to our own Saturn (0.83\,\RJ), which has an upper limit 
on its core mass estimated to be $20\rm\,M_{\oplus}$ (Guillot~\&\ Gautier~\cite{guillot14}).
Results from their interior modeling yield a core mass of $32 \pm 7 \rm\,M_{\oplus}$ and 
$33^{+7}_{-4}\rm\,M_{\oplus}$,  for KOI-188b and KOI-195b  and assuming dissipation. 
In fact, their estimated heavy elements content greatly depends on whether we dissipate stellar 
energy or not into the planets. With standard models (i.e., without dissipation), KOI-188b would 
have only $8^{+7}_{-4}\rm\,M_{\oplus}$, and KOI-195b just $2^{+5}_{-2}\rm\,M_{\oplus}$. 
Therefore, coreless models cannot be fully excluded from the possible solutions. 
However, many irradiated planets require a mechanism to dissipate external energy in their interior 
in order to explain their observed radius (Guillot~\&\ Gautier~\cite{guillot14}). Thus, solutions taking 
into account some dissipation are preferred and should provide an upper-limit for the core-mass 
values given here.
Interestingly, the uncertainties on the planetary mass of these two objects are large enough to 
allow for  sub-Uranus mass planets at a 3-$\sigma$ level. Our models can only go down as far as 0.04\,\RJ\ 
 in masses, yet it is enough to put a lower limit on the masses of KOI-188b and 
KOI-195b based on these models: with dissipation, both should have masses 
larger than $\sim0.1\rm\,M_{Jup}$ (2\,$\sigma$) in order to explain~their~measured radius and age, 
regardless of the amount of heavy elements we put in the core. However, with standard models 
we can put this same lower limit only on KOI-188b, if and only if we assume the planet has between 
about 5 and $10\rm\,M_{\oplus}$ of heavy~elements.

Lastly, KOI-188b orbits an over-metallic K1V star, whereas KOI-195b orbits an 
under-metallic~G1V~star, which are the two main differences between the two systems.
According to their metallicity (Guillot et al.~\cite{guillot06}), KOI-195b should 
likely have lighter core than KOI-188b. This would imply that, somehow, KOI-195b would 
dissipate the incoming stellar flux less effectively  than KOI-188b could, implying 
differences in their atmospheric properties.

The planet KOI-192b orbits a G2V star. 
It has similar mass and radius ($0.29 \pm 0.09$~\MJ\ and $1.23 \pm 0.21$~\RJ) to the two 
planets above, but its orbital period of $10.2909940 \pm 0.0000011$~days is longer. 
The measured mass agrees with the $0.6$\,\MJ\ upper limit obtained with SOPHIE by 
Santerne et al.~(\cite{santerne12}). 
Our HARPS-N data did not allow us to significantly detect an eccentricity.
We formally obtain $e=0.35\pm0.11$ 
(corresponding to an argument of the pericenter $\omega=93.6^{\circ}  \pm  0.5^{\circ}$),
but that value only stands on a few points, which could lead 
to a slight overestimation of the eccentricity in the case of correlated noise 
(see, e.g., Husnoo et al.~\cite{husnoo11}).
Thus, we adopt the conservative 99\,\%\ upper limit $e<0.57$. 
Eccentric orbits imply larger stellar and planetary radii than a circular orbit for KOI-192b, 
and lower corresponding densities.
The large uncertainty on the stellar density results in a large uncertainty on its precise 
evolutionary status. As a consequence, among our four planet host-stars, its radius and its 
inferred luminosity have the largest statistical uncertainties.
As seen on the upper panel of 
Fig.~\ref{fig_stat}, KOI-192b lies in a parameter space area 
(periods between 7 and 14~days and masses between 0.1 and 1.5~\MJ)
where only four other transiting planets are known, 
namely WASP-59b, CoRoT-4b, KELT-6b, and HAT-P-17b (H\'ebrard et al.~\cite{hebrard13b}; 
Moutou et al.~\cite{moutou08}; Collins et al.~\cite{collins14}; Howard et al.~\cite{howard12}), 
whereas \kepler\ allowed numerous planets of lower masses to be detected 
at these orbital periods (seen at the bottom of the upper panel of Fig.~\ref{fig_stat}).
These four planets are all more massive than KOI-192b. The first three  ones 
(WASP-59b, CoRoT-4b, and KELT-6b) have low-eccentricity orbits, 
whereas HAT-P-17b has an orbit with $e=0.342 \pm 0.006$ and a long-period, cold Jupiter companion.
The non-transiting planet HD\,108147b (Pepe et al~\cite{pepe02}; Butler et al.~\cite{butler06}) has  
similar period and (sky-projected) mass, but its radius is unknown; its eccentricity 
is $0.498 \pm 0.025$.
Additional observations are mandatory to constrain the eccentricity of KOI-192b.
Transiting planets in that long-period range are difficult to detect from 
ground-based photometry. The low mass also makes it  difficult to detect 
in radial velocity on faint host stars, such as those observed by Kepler 
and CoRoT. 
So the scarcity of transiting planets with such orbital periods and masses 
could partly be due to observational biases. Radial velocity surveys
detected a small amount of giant planets at intermediate orbital periods between 
about 10 and 100 days (the so-called period valley; see, e.g., Cumming et al.~\cite{cumming08};
Mayor et al.~\cite{mayor11}), but it is not as strong as the lack of giant planets on 
long periods seen in the upper panel of Fig. 4. Still, it remains difficult to 
distinguish true trends from observational biases.

We estimate the KOI-192b core-mass to be $16^{+22}_{-7}\rm\,M_{\oplus}$ 
with dissipation models, and $1^{+4}_{-1}\rm\,M_{\oplus}$ with standard ones. The larger relative 
error on these values are due to the larger error on the radius. This means that it is impossible to put 
any lower limit on the mass of KOI-192b based on the theoretical models we used.
Despite its longer orbital period, the equilibrium temperature of KOI-192b is similar to that of KOI-188b
because of the higher temperature of the host star. KOI-192b is cooler than the closer-in planet KOI-195b 
by about 200\,K, but it could have a larger radius, which is not the expected trend.

Finally, KOI-830b has a mass of $1.27 \pm 0.19$\,\MJ\ and a radius of $1.08 \pm 0.03$\,\RJ, 
corresponding to a density similar to that of Jupiter (lower panel of Fig.~\ref{fig_stat}). 
Its short period of $3.52563254 \pm 0.00000015$~days makes it a 
standard hot Jupiter, which orbits a K1V star which could be slightly over-metallic.
Its orbit is expected to be circularized, and we derived the 99\,\%\ upper limit $e<0.22$.
In its mass and radius range, there are no solutions from standard models, while dissipated-energy 
models provide a core-mass estimation of $45^{+25}_{-15}\rm\,M_{\oplus}$.
This is in line with what  would be expected for a hot Jupiter orbiting around a metal-rich star.
In comparison with the three other systems presented here, KOI-830 shows 
the larger disagreement between $(R_{\rm p}/R_{\star})^2$ and the observed depth of the transits.
The transit depth depends on $(R_{\rm p}/R_{\star})^2$, but also on other parameters, including 
the limb darkening coefficients and the impact parameter. 
We note that KOI-830 and KOI-188 have similar limb-darkening coefficients but 
distinct impact parameters.

As for most known transiting planets, these four systems are not Darwin stable, and the 
ultimate fate of these close-in planets is to spiral into their stars
(see, e.g., H\'ebrard et al.~\cite{hebrard13a}) although this evolution can 
be particularly slow for such low-mass planets. The characteristic timescales of tidal evolution 
depend on the tidal dissipation efficiency $Q'_\star$ of their host star (Matsumura et al.~\cite{matsumura10}). 
Even considering a relatively efficient dissipation of $Q'_\star=10^6$, only KOI-830 would risk 
being engulfed before the end of the host's main-sequence.

\section{Conclusions}
\label{sect_concle}

We have presented the four new transiting, giant planets
KOI-188b, KOI-195b, KOI-192b, and KOI-830b.
They were detected thanks to \kepler\ photometry 
complemented by HARPS-N spectroscopic~observations.
The joined analysis of the datasets allowed us to characterize the four  
systems. When compared to the parameters initially derived 
by the \kepler\ Team from the initial \kepler\ light curves only (Brown et al.~\cite{brown11}; 
Borucki et al.~\cite{borucki11a}, \cite{borucki11b}; Batalha et al.~\cite{batalha12}), 
our derived parameters agree but are more accurate and robust.
The identification of the planetary nature of the four objects as well as the planetary 
mass measurements are reported for the first time~here.
After the~announcement of the present results, the \kepler\ Team gave to systems 
KOI-188, KOI-195, KOI-192, and KOI-830 the names 
Kepler-425, Kepler-426, Kepler-427, and Kepler-428, respectively 
(see Table~\ref{startable_KOI}).

These new secured planets improve the statistics of well-described 
planetary systems, in particular in the  Saturn-mass regime where only few cases are known.
They also provide  new targets for follow-up studies on individual~systems.
Our observations also include a pre-screening with the SOPHIE spectrograph, which revealed 
KOI-219.01 to be a false positive. This confirms the fact that \kepler\ transiting candidates
include targets which are actually  not planets, in particular among the KOIs corresponding 
to close-in, giant companions. Follow-up studies are necessary both to establish the planetary 
nature of most of the transiting candidates identified from photometric surveys and to measure 
their mass.

\begin{acknowledgements}
This publication is based on observations collected with the NASA satellite \kepler, 
the HARPS-N spectrograph on the 3.58~m Italian \textit{Telescopio Nazionale Galileo}
(TNG) operated on the island of La Palma by the Fundaci\'on Galileo Galilei of the INAF 
(Instituto Nazionale di Astrofisica) at the Spanish Observatorio del Roque de los Muchachos 
of the Instituto de Astrofisica de Canarias
(programs OPT13A\_8 and OPT13B\_30
from OPTICON common time allocation process for EC 
supported trans-national access to European telescopes), 
and the SOPHIE spectrograph on the 1.93~m telescope at \textit{Observatoire de Haute-Provence} (CNRS), 
France (programs 11A.PNP.MOUT, 12A.PNP.MOUT, and 13A.PNP.MOUT).
The authors particularly thank the \kepler, TNG, and OHP teams, whose work and expertise allowed
these results to be obtained.
This research has made use of 
the Extrasolar Planets Encyclopaedia (exoplanet.eu)
and
the Exoplanet Orbit Database (exoplanets.org).
The research leading to these results has received funding from the 
``Programme National de Plan\'etologie'' (PNP) of CNRS/INSU, 
and from the
European Community's Seventh 
Framework Programme (FP7/2007-2013) under grant agreement number RG226604 (OPTICON).
AS acknowledges the support by the European Research Council/European Community under the 
FP7 through Starting Grant agreement number 239953.
AS is supported by the European Union under a Marie Curie Intra-European Fellowship for Career 
Development with reference FP7-PEOPLE-2013-IEF, number 627202.
ASB acknowledges funding from the European Union Seventh Framework Programme (FP7/2007-2013) 
under Grant agreement n. 313014 (ETAEARTH).
\end{acknowledgements}

\Online

\begin{table*}
\caption{(Electronic table available online.) Priors used for the analysis of KOI-188, KOI-195, KOI-192, and KOI-830.
$\mathcal{U}(a,b)$ denotes a Uniform prior between $a$ and $b$; $\mathcal{J}(a,b)$ denotes a Jeffreys 
distribution between $a$ and $b$; $\mathcal{N}(\mu,\sigma^{2})$ denotes a Normal distribution with a 
mean of $\mu$ and a width of $\sigma^{2}$; $\mathcal{N_{A}}(\mu,\sigma_{-}^{2}, \sigma_{+}^{2})$ 
denotes an asymmetric Normal distribution with mean $\mu$, upper width $\sigma_{+}^{2}$, and lower 
width $\sigma_{-}^{2}$; $\mathcal{N_{U}}(\mu,\sigma^{2}, a, b)$ denotes a Normal distribution with a 
mean of $\mu$, a width of $\sigma^{2}$, and limited by a Uniform distribution between $a$ and $b$; 
$\mathcal{N}_{2}(\mu_{1},\sigma_{1}^{2}, \mu_{2},\sigma_{2}^{2}, \Delta_{A})$ denotes a Bi-Normal 
distribution with means of $\mu_{1}$ and $\mu_{2}$, widths of $\sigma_{1}^{2}$ and $\sigma_{2}^{2}$,  
and $\Delta_{A}$ is the amplitude ratio between the two Normal distributions such that 
$\mathcal{N}_{2}(\mu_{1},\sigma_{1}^{2}, \mu_{2},\sigma_{2}^{2}, 
\Delta_{A}) = 0.5\times[\mathcal{N}(\mu_{1},\sigma_{1}^{2})\times(1-\Delta_{A}) + 
\mathcal{N}(\mu_{2},\sigma_{2}^{2})\times(1+\Delta_{A})]$; $\mathcal{S}(a,b)$ denotes 
a Sine distribution between $a$ and $b$; and finally $\beta(a,b)$ denotes a Beta distribution 
with parameters $a$ and $b$ (Kipping~\cite{kipping13}).}
\begin{tabular}{lccccc}
\hline
Parameter & KOI-188 & KOI-195 & KOI-192 & KOI-830 \\
\hline
\multicolumn{5}{l}{\hspace{-0.3cm} \it Ephemerides and orbital parameters}\\
Orbital period $P$ [d] & $\mathcal{N}$(3.79702, 5 10$^{-5}$) & $\mathcal{N}$(3.21752, 5 10$^{-5}$) & $\mathcal{N}$(10.29100, 5 10$^{-5}$) & $\mathcal{N}$(3.52563, 5 10$^{-5}$)  \\
Transit epoch $T_{0}$ [BJD$_\mathrm{TDB}$ - 2454900] & $\mathcal{N}(66.50811, 0.002)$ & $\mathcal{N}(66.63102, 0.002)$ &  $\mathcal{N}(70.02113, 0.002)$ & $\mathcal{N}(103.04727, 0.002)$ \\
Orbital inclination $i$ [$^{\circ}$] & $\mathcal{S}(80,90)$ & $\mathcal{S}(80,90)$ & $\mathcal{S}(80,90)$ & $\mathcal{S}(80,90)$ \\
Orbital eccentricity $e$ & 0 & 0 & $\beta(0.867,3.03)$ & 0  \\
Argument of periastron $\omega$ [$^{\circ}$] & 90 & 90 & $\mathcal{U}(0,360)$ & 90 \\
\hline
\multicolumn{5}{l}{\hspace{-0.3cm} \it Transit parameters}\\
Radius ratio $r_{p}/R_{\star}$ & $\mathcal{J}(0.01, 0.5)$ & $\mathcal{J}(0.01, 0.5)$ & $\mathcal{J}(0.01, 0.5)$ & $\mathcal{J}(0.01, 0.5)$ \\
Linear limb-darkening coefficient $u_{a}$ & $\mathcal{U}(-0.5,1.2)$ & $\mathcal{U}(-0.5,1.2)$ & $\mathcal{U}(-0.5,1.2)$ & $\mathcal{U}(-0.5,1.2)$  \\
Quadratic limb-darkening coefficient $u_{b}$ & $\mathcal{U}(-0.5,1.2)$ & $\mathcal{U}(-0.5,1.2)$ & $\mathcal{U}(-0.5,1.2)$ & $\mathcal{U}(-0.5,1.2)$ \\
\hline
\multicolumn{5}{l}{\hspace{-0.3cm} \it Radial velocity parameters}\\
Systemic velocity $\upsilon_{0}$ [km\,s$^{-1}$] & $\mathcal{U}(-50,-40)$ & $\mathcal{U}(-85,-70)$ & $\mathcal{U}(-30,-20)$ & $\mathcal{U}(-30,-15)$ \\ 
Radial-velocity semi-amplitude $K$ [km\,s$^{-1}$] & $\mathcal{U}(0, 1)$ & $\mathcal{U}(0, 1)$ & $\mathcal{U}(0, 1)$ & $\mathcal{U}(0, 1)$ \\
\hline
\multicolumn{5}{l}{\hspace{-0.3cm} \it Stellar parameters}\\
Effective temperature \teff\, [K] & $\mathcal{N}(5110, 100)$ & $\mathcal{N}(5710, 100)$ & $\mathcal{N}(5740, 90)$ & $\mathcal{N}(5160, 200)$ \\
Iron abundance \met\, [dex] & $\mathcal{N}(0.35, 0.13)$ & $\mathcal{N}(-0.19, 0.08)$ & $\mathcal{N}(-0.19, 0.07)$ & $\mathcal{N}(0.23, 0.29)$\\
Bulk density $\rho_{\star}$ [$\rho_{\odot}$] & $\mathcal{N_{A}}$(1.19, 0.22, 0.17) & $\mathcal{N}$(1.13, 0.26) & $\mathcal{N}_{2}$(0.27, 0.09, & $\mathcal{N}$(1.46, 0.31) \\
 &  &  & 0.66, 0.33, -0.56)  &  \\
\hline
\multicolumn{5}{l}{\hspace{-0.3cm} \it System parameters}\\
Distance from Earth $D$ [pc] & $\mathcal{U}(10, 5000)$ & $\mathcal{U}(10, 5000)$ & $\mathcal{U}(10, 5000)$ & $\mathcal{U}(10, 5000)$ \\
Interstellar extinction $E(B-V)$  [mag] & $\mathcal{U}(0,2)$ & $\mathcal{U}(0,2)$ & $\mathcal{U}(0,2)$ & $\mathcal{U}(0,2)$ \\
\hline
\multicolumn{5}{l}{\hspace{-0.3cm} \it Instrumental parameters}\\
\multicolumn{5}{l}{\hspace{0.5cm} \it \textit{Kepler} season 0 LC: }\\
Jitter [\%] & $\mathcal{U}(0, 1)$ & $\mathcal{U}(0, 1)$ & $\mathcal{U}(0, 1)$ & $\mathcal{U}(0, 1)$ \\
Contamination [\%] & $\mathcal{N_{U}}(4.8, 1, 0, 100)$ & $\mathcal{N_{U}}(3.7, 1, 0, 100)$ & $\mathcal{N_{U}}(4.0, 1, 0, 100)$ & $\mathcal{N_{U}}(6.6, 1, 0, 100)$ \\
Out-of-transit flux & $\mathcal{U}(0.999, 1.001)$ & $\mathcal{U}(0.999, 1.001)$ & $\mathcal{U}(0.999, 1.001)$ & $\mathcal{U}(0.999, 1.001)$ \\
\multicolumn{5}{l}{\hspace{0.5cm} \it \textit{Kepler} season 0 SC: }\\
Jitter [\%] & $\mathcal{U}(0, 1)$ & $\mathcal{U}(0, 1)$ & $\mathcal{U}(0, 1)$ & -- \\
Contamination [\%] & $\mathcal{N_{U}}(4.8, 1, 0, 100)$ & $\mathcal{N_{U}}(3.7, 1, 0, 100)$ & $\mathcal{N_{U}}(4.0, 1, 0, 100)$ & -- \\
Out-of-transit flux & $\mathcal{U}(0.999, 1.001)$ & $\mathcal{U}(0.999, 1.001)$ & $\mathcal{U}(0.999, 1.001)$ & -- \\
\multicolumn{5}{l}{\hspace{0.5cm} \it \textit{Kepler} season 1 LC:}\\
Jitter [\%] & $\mathcal{U}(0, 1)$ & $\mathcal{U}(0, 1)$ & $\mathcal{U}(0, 1)$ & $\mathcal{U}(0, 1)$ \\
Contamination [\%] & $\mathcal{N_{U}}(3.7, 1, 0, 100)$ & $\mathcal{N_{U}}(4.1, 1, 0, 100)$ & $\mathcal{N_{U}}(3.2, 1, 0, 100)$ & $\mathcal{N_{U}}(5.4, 1, 0, 100)$ \\
Out-of-transit flux & $\mathcal{U}(0.999, 1.001)$ & $\mathcal{U}(0.999, 1.001)$ & $\mathcal{U}(0.999, 1.001)$ & $\mathcal{U}(0.999, 1.001)$ \\
\multicolumn{5}{l}{\hspace{0.5cm} \it \textit{Kepler} season 1 SC:}\\
Jitter [\%] & $\mathcal{U}(0, 1)$ & -- & $\mathcal{U}(0, 1)$ & -- \\
Contamination [\%] & $\mathcal{N_{U}}(3.7, 1, 0, 100)$ & -- & $\mathcal{N_{U}}(3.2, 1, 0, 100)$ & -- \\
Out-of-transit flux & $\mathcal{U}(0.999, 1.001)$ & -- & $\mathcal{U}(0.999, 1.001)$ & -- \\
\multicolumn{5}{l}{\hspace{0.5cm} \it \textit{Kepler} season 2 LC:}\\
Jitter [\%] & $\mathcal{U}(0, 1)$ & $\mathcal{U}(0, 1)$ & $\mathcal{U}(0, 1)$ & $\mathcal{U}(0, 1)$ \\
Contamination [\%] & $\mathcal{N_{U}}(5.8, 1, 0, 100)$ & $\mathcal{N_{U}}(8.6, 1, 0, 100)$ & $\mathcal{N_{U}}(3.1, 1, 0, 100)$ & $\mathcal{N_{U}}(8.7, 1, 0, 100)$ \\
Out-of-transit flux & $\mathcal{U}(0.999, 1.001)$ & $\mathcal{U}(0.999, 1.001)$ & $\mathcal{U}(0.999, 1.001)$ & $\mathcal{U}(0.999, 1.001)$ \\
\multicolumn{5}{l}{\hspace{0.5cm} \it \textit{Kepler} season 2 SC:}\\
Jitter [\%] & $\mathcal{U}(0, 1)$ & $\mathcal{U}(0, 1)$ & $\mathcal{U}(0, 1)$ & -- \\
Contamination [\%] & $\mathcal{N_{U}}(5.8, 1, 0, 100)$ & $\mathcal{N_{U}}(8.6, 1, 0, 100)$ & $\mathcal{N_{U}}(3.1, 1, 0, 100)$ & -- \\
Out-of-transit flux & $\mathcal{U}(0.999, 1.001)$ & $\mathcal{U}(0.999, 1.001)$ & $\mathcal{U}(0.999, 1.001)$ & -- \\
\multicolumn{5}{l}{\hspace{0.5cm} \it \textit{Kepler} season 3 LC:}\\
Jitter [\%] & $\mathcal{U}(0, 1)$ & $\mathcal{U}(0, 1)$ & $\mathcal{U}(0, 1)$ & $\mathcal{U}(0, 1)$ \\
Contamination [\%] & $\mathcal{N_{U}}(3.3, 1, 0, 100)$ & $\mathcal{N_{U}}(7.1, 1, 0, 100)$ & $\mathcal{N_{U}}(3.5, 1, 0, 100)$ & $\mathcal{N_{U}}(8.9, 1, 0, 100)$ \\
Out-of-transit flux & $\mathcal{U}(0.999, 1.001)$ & $\mathcal{U}(0.999, 1.001)$ & $\mathcal{U}(0.999, 1.001)$ & $\mathcal{U}(0.999, 1.001)$ \\
\multicolumn{5}{l}{\hspace{0.5cm} \it \textit{Kepler} season 3 SC:}\\
Jitter [\%] & $\mathcal{U}(0, 1)$ & $\mathcal{U}(0, 1)$ & $\mathcal{U}(0, 1)$ & -- \\
Contamination [\%] & $\mathcal{N_{U}}(3.3, 1, 0, 100)$ & $\mathcal{N_{U}}(7.1, 1, 0, 100)$ & $\mathcal{N_{U}}(3.5, 1, 0, 100)$ & -- \\
Out-of-transit flux & $\mathcal{U}(0.999, 1.001)$ & $\mathcal{U}(0.999, 1.001)$ & $\mathcal{U}(0.999, 1.001)$ & -- \\
\multicolumn{5}{l}{\hspace{0.5cm}  \it HARPS-N 2012:}\\
Jitter [\ms] & $\mathcal{U}(0, 100)$ & $\mathcal{U}(0, 100)$ & $\mathcal{U}(0, 100)$ & -- \\
Offset [\ms] & $\mathcal{U}(-200, 200)$ & $\mathcal{U}(-200, 200)$ & $\mathcal{U}(-200, 200)$  & $\mathcal{U}(-200, 200)$ \\
\multicolumn{5}{l}{\hspace{0.5cm} \it HARPS-N 2013:}\\
Jitter [\ms] & $\mathcal{U}(0, 100)$ & $\mathcal{U}(0, 100)$ & $\mathcal{U}(0, 100)$ & $\mathcal{U}(0, 100)$  \\
Offset [\ms] & 0 & 0 & 0 & 0 \\
\multicolumn{5}{l}{\hspace{0.5cm} \it SED:}\\
Jitter [mag] & $\mathcal{U}(0,1)$ & $\mathcal{U}(0,1)$ & $\mathcal{U}(0,1)$ & $\mathcal{U}(0,1)$  \\
\hline
\label{PriorTable}
\end{tabular}
\end{table*}

\end{document}